\documentclass[LectureNotes]{SciPost}

\usepackage{hyperref}
\hypersetup{
    bookmarks=true,         % show bookmarks bar?
    unicode=false,          % non-Latin characters in Acrobat�s bookmarks
    pdftoolbar=true,        % show Acrobat�s toolbar?
    pdfmenubar=true,        % show Acrobat�s menu?
    pdffitwindow=false,     % window fit to page when opened
    pdfstartview={FitH},    % fits the width of the page to the window
    pdftitle={Lecture notes on Many-Body-Localization},    % title
    pdfauthor={},     % author
    pdfsubject={},   % subject of the document
    pdfcreator={},   % creator of the document
    pdfproducer={}, % producer of the document
    pdfkeywords={keyword1} {key2} {key3}, % list of keywords
    pdfnewwindow=true,      % links in new window
    colorlinks=true,       % false: boxed links; true: colored links
    linkcolor=[rgb]{0.55,0,0.5},          % color of internal links (change box color with linkbordercolor)
    citecolor=[rgb]{0,0,0.6},        % color of links to bibliography
    filecolor=magenta,      % color of file links
    urlcolor=blue           % color of external links
}
\usepackage{graphicx}

\usepackage{braket}
\usepackage[makeroom]{cancel}

\newcommand{\bea}{\begin{eqnarray}}
\newcommand{\eea}{\end{eqnarray}}
\newcommand{\be}{\begin{equation}}
\newcommand{\ee}{\end{equation}}
\newcommand{\nn}{{\nonumber}}

\newcommand{\JZ}{{\mathbb{Z}}}
\newcommand{\JN}{{\mathbb{N}}}
\newcommand{\JR}{{\mathbb{R}}}
\newcommand{\JP}{{\mathbb{P}}}
\newcommand{\JC}{{\mathbb{C}}}

\newcommand{\cT}{{\cal T}}
\newcommand{\cO}{{\cal O}}
\newcommand{\cI}{{\cal I}}
\newcommand{\cJ}{{\cal J}}
\newcommand{\cA}{{\cal A}}
\newcommand{\cE}{{\cal E}}
\newcommand{\cP}{{\cal P}}
\newcommand{\cD}{{\cal D}}
\newcommand{\cchi}{{\cal \chi}}

\begin{document}

% TODO: write your article's title here. 
% The article title is centered, Large boldface, and should fit in two lines
\begin{center}{\Large \textbf{Perturbation theory approaches to Anderson and Many-Body Localization: some lecture notes
}}\end{center}

% TODO: write the author list here. Use initials + surname format.
% Separate subsequent authors by a comma, omit comma at the end of the list.
% Mark the corresponding author with a superscript *. 
\begin{center}
A. Scardicchio\textsuperscript{1,2}, T. Thiery \textsuperscript{3*}
\end{center}

% TODO: write all affiliations here. 
% Format: institute, city, country
\begin{center}
{\bf 1} Abdus Salam ICTP, Strada Costiera 11, 34151 Trieste, Italy \\
{\bf 2} INFN, Sezione di Trieste, Via Valerio 2, 34127 Trieste, Italy
\\
{\bf 3} Instituut voor Theoretische Fysica, KU Leuven, 3001 Leuven, Belgium.
\\
% TODO: provide email address of corresponding author
* thimothee.thiery@kuleuven.be
\end{center}

% For convenience during refereeing: line numbers
%\linenumbers

\section*{Abstract}
{\bf 
These are lecture notes based on three lectures given by Antonello Scardicchio at the December 2016 Topical School on Many-Body-Localization organized by the Statistical Physics Group of the Institute Jean Lamour in Nancy. They were compiled and put in a coherent logical form by Thimoth\'ee Thiery.
}

% TODO: include a table of contents (optional)
% Guideline: if your paper is longer that 6 pages, include a TOC
% To remove the TOC, simply cut the following block
\vspace{10pt}
\noindent\rule{\textwidth}{1pt}
\tableofcontents\thispagestyle{fancy}
\noindent\rule{\textwidth}{1pt}
\vspace{10pt}

\section{Introduction}
Anderson localization \cite{Anderson1958} is the suppression of diffusion for a particle in a disordered potential due to quantum effects. In a nutshell, quantum mechanics typically dictates a discrete spectrum for the forces (intended as the expectation value of the acceleration divided by the mass) acting on a test particle. Think of a harmonic oscillator or the bound states of the hydrogen atom. Diffusive forces (Langevin forces), on the other hand, necessitate a continuous spectrum (not necessarily white) and therefore the phenomenology of quantum mechanics leaves the door open to absence of diffusion.

To give an example of this phenomenon, assume a particle can jump with amplitude $g$ between a central site 0 with energy 0 and $2d$ sites with energy $\epsilon_i$ uniformly distributed in an interval of size $W$. Diagonalizing the $(2d+1)\times(2d+1)$ Hamiltonian we obtain $2d+1$ discrete eigenvalues. In absence of any symmetry the evolution of the wave function will be quasiperiodic, returning with high probability (say $>1/2$) to the central site 0 in any sufficiently large interval of time. The local density of states 
\begin{equation}
\rho(0,E)=|\braket{0|E}|^2
\end{equation}
is the Fourier spectrum of the wave function $\psi(0,t)=\bra{0}e^{-iHt}\ket{0}$ and it is a \emph{pure point} distribution, in the language of spectral theory, with support on $2d+1$ energies which are the eigenvalues. It is a distribution, in the sense that it is positive and of finite (unit) mass.

As the arrangement of sites expands to become an infinite lattice (the thermodynamic limit) we envision two possibilities: either the motion becomes chaotic, and the return probability goes to zero, or the wave function retains a quasi-periodic behavior and the return probability becomes very often large (it stays $O(1)$ instead of $O(1/volume)$, namely there is an $\epsilon>0$ such that in the thermodynamic limit the return probability is $>\epsilon$). The first case corresponds to the local density of states being a \emph{continuous} measure $\rho(0,E)$ and the second to \emph{pure point} measure.
 
The question is then: What happens when particles interact? Intuitively localization was caused by the mismatch between the energies on different sites $\epsilon_i$ not being compensated by the kinetic energy of $O(g)$. Interaction then should destabilize localization, as now we can borrow (or give) energy from other particles to compensate the energy differences of the sites. One could even push this argument to argue that interaction must destroy localization (as was one prominent line of thought before Basko-Aleiner-Altshuler (BAA) breakthrough \cite{BAA2006}). While this is not the case (with a notable exception being long-range interactions, \cite{yao2014many} but see also \cite{nandkishore2017many,brenes2017many}), interaction certainly restricts the region of existence of localization in parameter space.

It turns out that one can find a stable localized region, if the interaction is sufficiently weak (with respect to hopping $g$) and the disorder is sufficiently strong (again with respect to $g$). In order to prove it, it is convenient to work out the perturbation theory for weak interaction, after having assumed that the disorder is sufficiently strong to localize all the non-interacting states. This expansion is similar to the ``locator expansion" of Anderson and therefore, pedagogically, we thought to be a good idea to start with it and go in some detail in its analysis.

In general, in these lectures (originally this material was covered in three lectures) we will skip a description of the phenomenology of MBL (although some of it will emerge naturally from the analytic results) which was covered in other lectures and other topics which were quite instrumental in understanding of the properties of this new phase (like the behavior of entanglement). We refer the interested reader to recent reviews \cite{NandkishoreHuseReview2015,AltmanVoskReview2015,ImbrieRosScardicchio2016review} on the topic.

The outline of the lecture notes is as follows. Sec.~\ref{sec:AL1} is an introduction to Anderson localization (AL). Sec.~\ref{sec:AL2} gives two analytical approaches to AL: the self-consistent approximation of \cite{AbouChacraThoulessAnderson1973a} and the forward approximation. Sec.~\ref{sec:MBL1} is an introduction to Many-Body-Localization (MBL). Sec.~\ref{sec:MBL2} reviews the forward approximation approach to MBL originally developed in \cite{RosMullerScardicchio2015}.

\section{Anderson Localization} \label{sec:AL1}

\subsection{Introduction} \label{subsec:AL1:Intro}

In the seminal paper \cite{Anderson1958}, Anderson analyzed the problem of the propagation of a quantum particle in a disordered potential under unitary time evolution. The model that he considered is now known as the Anderson model and is defined as follows:

\paragraph{Definition} Considering the lattice $\Lambda:=\JZ^d$ with sites labeled by $i \in \JZ^d$, the Anderson Hamiltonian is
\bea \label{Eq:AndersonHamiltonian}
H := - g  \left( \sum_{ \langle i j \rangle} c_i^{\dagger} c_j  + c_j^{\dagger} c_i \right) + \sum_{i } \epsilon_i c_i^{\dagger} c_i \, ,
\eea
where: (i) the operators $(c_i , c_i^{\dagger})$ denote the fermionic creation/annihilation operators at site $i \in \JZ^d$ satisfying the algebra $\{c_i^\dagger , c_j \} = \delta_{ij}$ and  $\{c_i^\dagger , c_j^\dagger \} =\{c_i , c_j \}  =0  $; (ii) $g \geq 0$ is the hopping amplitude and the sum $\sum_{\langle ij \rangle}$ denotes the sum over nearest neighbour sites on the lattice; (iii) the on-site energies $\epsilon_i$ are independent, identically distributed (iid) random variables (RVs) that are uniformly distributed in $[-W/2,W/2]$, with $W$ the amplitude of the `disorder'. $H$ acts on the Fock space ${\cal H}$ that is constructed from the fermionic operators: ${\cal H} := {\rm Vect}\left( \otimes_{i \in \JZ^d}  \ket{\eta_i}  , \eta_j \in \{0 ,1\}  \right)$ with $c_i \ket{\eta }_i  = \delta_{\eta , 1} \ket{0}_i$ and $c_i^{\dagger} \ket{\eta }_i  = \delta_{\eta , 0} \ket{1}_i$. Since $H$ commutes with $N := \sum_{i} c_i^{\dagger} c_i$, it conserves the number of Fermions, and since it is quadratic in the fermionic operators it is a model for the propagation of {\it independent} electrons in a random short-range correlated potential. We can thus restrict our analysis to the one-fermion subspace ${\cal H}_1$, where for $n \in \JN$ we denote ${\cal H}_n := {\rm Vect}\left( \otimes_{i \in \JZ^d}  \ket{\eta_i}_i  , \eta_j \in \{0 ,1\}   , \sum_j \eta_j = n \right) $. Using $\ket{i}$ as a shorthand for $\otimes_{j\neq i } \ket{0}_j \otimes \ket{1}_i$, the Anderson Hamiltonian restricted to ${\cal H}_1$ may be rewritten
\bea \label{Eq:AndersonHamiltonian2}
H|_{{\cal H}_1} = - g  \left( \sum_{ \langle i j \rangle}  \ket{j} \bra{i}+ \ket{i} \bra{j}  \right)  + \sum_{i } \epsilon_i \ket{i} \bra{i} \, .
\eea
And for any state $\ket{\psi} \in {\cal H}_1$ we may introduce the wave-function $\psi(i) := \braket{i | \psi}$. The action of $H$ on the wave-function is thus
\bea
(H|_{{\cal H}_1} \psi)(i)= - g  \sum_{\langle i j \rangle}  \psi(j) + \epsilon_i \psi(i) \, . 
\eea
This also identifies ${\cal H}_1$ with square summable functions: ${\cal H}_1 \simeq \ell^2(\Lambda, \JC)$. Finally, the problem initially considered by Anderson is to compute the return probability $p(t)$ for a particle initially localized around the origin:
\bea
p(t) = |\psi(t,0)|^2 \quad , \quad i \partial_t \psi(t,i) = (H \psi(t,.))(i) \quad , \quad \psi(t=0,i) = \delta_{i,0} \, .
\eea
The model can be straightforwardly generalized to other lattices and other distributions. In particular it will often be convenient to consider a regularized version of the model on the lattice $\Lambda_L := \JZ^d \cap [0,L]^d$, with periodic boundary conditions being assumed. This is because ${\cal H}_1$ is then a finite dimensional Hilbert space of dimension $d_{{\cal H}_{1}}=L^d$, and thus $H|_{{\cal H}_1}$ admits a complete set of eigenfunctions/eigenvectors. The situation for $L= +\infty$ is more subtle.
\medskip

We begin by treating carefully the two trivial extreme cases: the free case $W=0$ and the infinite disorder case $g=0$. We take care of first considering the model on the finite lattice $\Lambda_L $ to highlight the differences between the two cases that appear in the thermodynamic limit $L \to \infty$.

\begin{figure}
\centerline{\includegraphics[width=12cm]{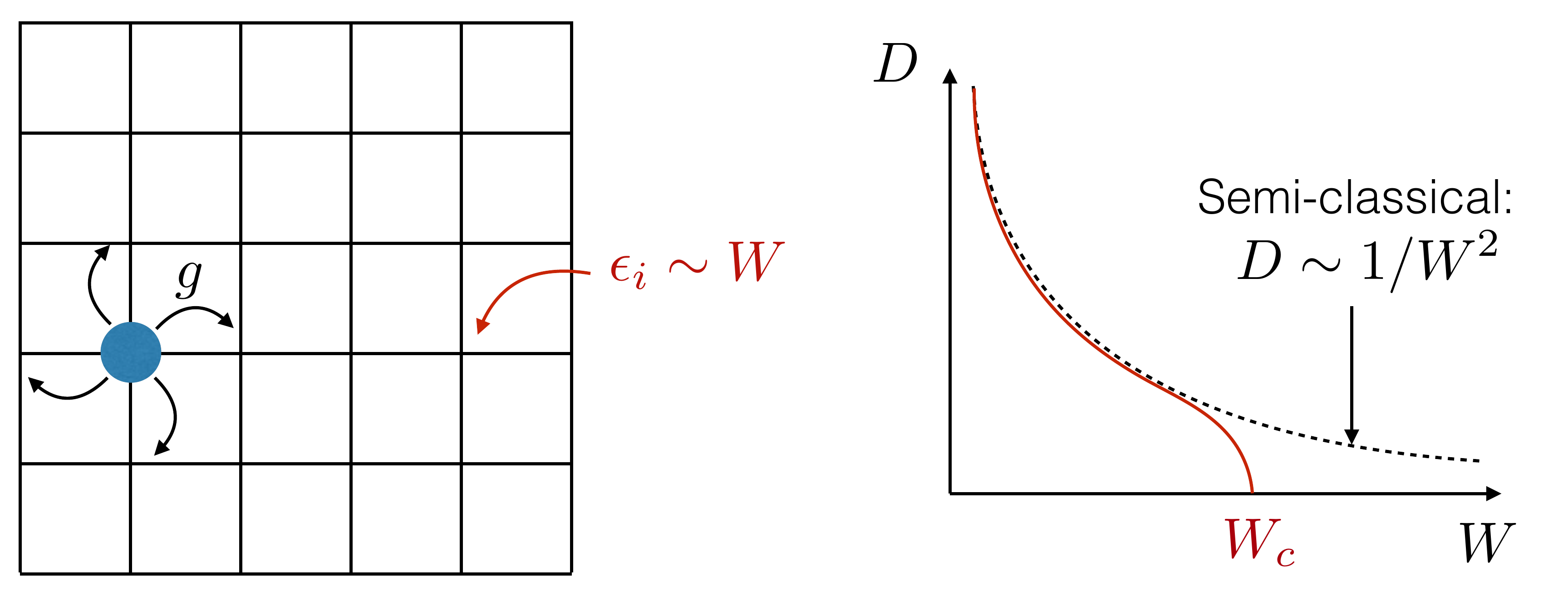}} 
\caption{The Anderson model and the localization transition: the diffusion coefficient of a quantum particle in a random potential vanishes with probability $1$ at a finite disorder strength.}
\label{Fig:AndersonModel}
\end{figure}

\subsubsection{Disorder Free case} \label{subsubsec:AL1:Intro:DisorderFree}

\paragraph{On a finite lattice}
In the absence of disorder $W=0$ and on the lattice $\Lambda_L $, the Hamiltonian is translationally invariant and its eigenstates/eigenfunctions defined by $H \psi_k = E_k \psi_k$ are the plane-waves
\bea
&& \psi_k(j) := \frac{1}{\sqrt{L^d}} e^{i k j} \quad , \quad E_k = - 2 g \sum_{i=1}^d \cos(k_i) \nn \\
&&  k = \frac{2\pi}{L}(n_1, \cdots,n_d)  \quad , \quad   (n_1, \cdots,n_d) \in \{0, \cdots , L-1\}^d  \nn \, .
\eea
The solution of the Anderson problem is thus
\bea
\psi(t,j) = \frac{1}{L^d} \sum_{k} e^{ i k  j + 2 i t g \sum_{i=1}^d \cos(k_i)} \nn \, .
\eea
In particular, the solution is periodic with a period that scales as $L$. This is due to quantum revivals and, although the wave-function propagates, it comes back to the origin an infinite number of times.

\paragraph{In the thermodynamic limit}

Taking $L \to \infty$ limit with $t$ fixed, the solution converges to 
\bea \label{Eq:solutionfreeevolutionBessel}
\psi(t,j) =  \int_{k \in [0,2\pi]^d} \frac{d^dk}{(2\pi)^d} e^{ i k  j + 2 i t g \sum_{i=1}^d \cos(k_i)} = i^d\prod_{i=1}^d  J_{j_i} (2 t g) \, ,
\eea
with $J_n(k)$ a Bessel function of the first kind. From the integral representation one immediately obtains that the return probability decays to $0$: $\psi(t,0)=\int_{k \in [-\pi,\pi]^d} \frac{d^dk}{(2\pi)^d} e^{- 2 i t g \sum_{i=1}^d \cos(k_i)}  \simeq \frac{e^{-2 i d t g}}{t^{d/2}} \int_{p \in \JR^d} \frac{d^dp}{(2\pi)^d} e^{- i g p^2}$ and
\bea
p(t) = |\psi(t,0) |^2  \sim 1/t^d \, .
\eea
While the center of the wave-packet remains around $0$, $x_t:= \sum_{j \in \JZ^d} j |\psi(t,j)|^2 = 0$, the mean-square displacement diverges balistically as
\bea  \label{Eq:dicreteballistictransport}
\Delta x_t^2 := \sum_{j \in \JZ^d} j^2 |\psi(t,j)|^2 = 2 d  g^2 t^2  \,.
\eea
The reason being that excitation propagates balistically in this system, and that the initial wave-packet is a superposition of plane-waves propagating in every directions. This ballistic transport is characteristic of a system that is invariant by translation, and thus conserves momentum. To see \eqref{Eq:dicreteballistictransport} note first from the representation in terms of Bessel functions (or integrals) and the normalization of the wave-function that $\Delta x_t^2(d) =   \sum_{i=1}^d \sum_{j_1 \in \JZ} \cdots \sum_{j_d \in \JZ} j_i^2 \prod_{k=1}^d |J_{j_k}(2tg)^2|  = d \sum_{j \in \JZ} j^2 |J_{j}(2tg)^2| =  d \Delta x_t^2(d=1)$. In $d=1$ we use that by definition $\Delta x_t^2 = \braket{\psi(t,.)|X^2|\psi(t,.)}$ with the position operator $X$ defined as $(X \phi)(j)=j\phi(j)$. Hence $\partial_t \Delta x_t^2  = i\braket{\psi(t,.)|[H,X^2]|\psi(t,.)}$, $\partial_t^2 \Delta x_t^2  = -\braket{\psi(t,.)|[H,[H,X^2]]|\psi(t,.)}$ and $\partial_t^3 \Delta x_t^2  = -i\braket{\psi(t,.)|[H,[H,[H,X^2]]]|\psi(t,.)}$. A direct calculation gives $([H,X^2]\phi)(j) = -g (\phi(j+1)+\phi(j-1) + 2 j (\phi(j+1)-\phi(j-1)))  = (H \phi)(j)  - 2 g j(\phi(j+1)-\phi(j-1))$ and $([H,[H,X^2]]\phi)(j) = -g( -2g \phi(j+2) + 4 g \phi(j) - 2 g \phi(j-2)) = 2(H^2 \phi)(j)$. Hence $\partial_t^3 \Delta x_t^2=0$ and $\Delta x_t^2 =  \Delta x_0^2 + t \braket{ \psi(t=0,.) |[H,X^2]|\psi(t=0,.)}+\frac{t^2}{2}\braket{\psi(t=0,.)|[H,[H,X^2]]|\psi(t=0,.)}$. Computing explicitly the different terms at $t=0$ leads to \eqref{Eq:dicreteballistictransport}.

\paragraph{Remark: no eigenstate in the thermodynamic limit} Let us finally note that, in the thermodynamic limit, the Hamiltonian $H$ has absolutely continuous spectrum $\sigma(H) = [-2dg,2dg]$ and therefore does not admit any eigenvectors. The eigenvectors on the finite lattice converge uniformly to $0$ in the thermodynamic limit: $|| \psi_k ||_{\infty} \sim 1/L^{d/2} \to_{L \to \infty} 0$. Although plane-waves are still formally eigenvectors of $H$, they are not normalizable for $L = \infty$, and are thus not part of ${\cal H}_1 \equiv \ell^2(\JZ^d , \JC)$. This goes along with the following behavior of the {\it local density of states}:
\bea
\rho(j,E) = \sum_{k} \delta(E-E_k) |\braket{k|j}|^2 \sim_{L \to \infty}  \int_{k \in [0,2\pi]^d} \frac{d^dk}{(2\pi)^d}  \delta(E + 2  g \sum_{i=1}^d \cos(k_i) ), \nn 
\eea
independent of $j$. In particular for $d=1$
\bea
\rho(j,E) =\frac{1}{2 \pi}  \frac{\theta(2g -|E|)}{\sqrt{4 g^2 -E^2}}.
\eea
Hence, in the thermodynamic limit, the local density of state defines a continuous distribution on the spectrum of $H$, $\sigma(H) = [-2dg,2dg]$. This is an example of \emph{absolutely continuous} spectrum (see below).

\subsubsection{Infinite disorder case} \label{subsubsec:AL1:Intro:InfiniteDisorder}

The small hopping/infinite disorder case $g/W=0$ is simpler, and nothing special occurs in the $L \to \infty$ limit. Indeed, the eigenvectors of $H$ on both $\Lambda$ and $\Lambda_L$ are simply given by $H \psi_i = E_i \psi_i$ with
\bea
\psi_i(j) = \delta_{j,i} \quad , \quad E_i = \epsilon_i \quad , \quad i \in \Lambda \, .  \nn 
\eea
The Anderson problem is trivial in this limit: $\psi(t,j) = e^{-i \epsilon_0 t} \delta_{j,0}$ and we have
\bea
p(t) = 1 \quad , \quad \Delta x_t^2 = 0 \, . 
\eea
This is an extreme example of localization: an initially localized wave-packet remains localized ad infinitum. The wave-packet does not spread and the return probability of the particle remains constant. Note finally that the local density of state trivially remains a discrete distribution, corresponding to a \emph{pure point} spectrum:
\bea
\forall L:\quad \rho(j,E) = \delta(E-\epsilon_j) \,. 
\eea
Pure point and continuous spectrum are the only two possibilities for spectra of Hamiltonians. The continuous spectrum can be further distinguished in absolutely and singular continuous.

\subsubsection{In between: diffusive motion} \label{subsubsec:AL1:Intro:Inbetween}

The difficult question is therefore to understand the case in between where there is a competition between the hopping term and the `localizing' disordered term of the Hamiltonian. Thinking of the evolution of the quantum particle from a classical perspective, one expects that the motion of the particle consists in a succession of ballistic propagation (in the region without disorder) and elastic scattering against the medium impurities. Such a picture irremediably leads to a diffusive propagation with
\bea
\Delta x_t^2 \simeq D t \, \quad \mbox{for}\quad t\gg \tau, 
\eea
with $D=v^2 \tau >0$ the diffusion coefficient, $v$ the typical velocity of the particle in between scattering events and $\tau$ the typical time between scattering over the impurities. Diffusion also implies a decay of the return probability as
\bea
p(t) \sim \frac{1}{(Dt)^{d/2}} \, .
\eea

The scattering rate can be found by adding disorder perturbatively on the clean, plane wave case. In the semiclassical case we can assume the particle has a definite momentum $k$ and compute the scattering probability per unit time. Since the spectrum is continuous one can apply the Fermi golden rule to obtain the scattering rate:
\bea
\frac{1}{\tau}=\Gamma_k&=&\pi \sum_{k'}\left|\bra{k}V\ket{k'}\right|^2\delta(E_k-E_{k'}) \nn \\ 
&=&\pi \sum_{k'} \left|\sum_j\epsilon_j\braket{k|j}\braket{j|k'}\right|^2\delta(E_k-E_{k'}) \nn \\ 
&=&\pi \sum_{k'} \left(\sum_j\epsilon_j^2\frac{1}{L^{2d}}+\sum_{j\neq j'}\epsilon_{j}\epsilon_{j'}\frac{e^{i(k-k')(j-j')}}{L^{2d}}\right)\delta(E_k-E_{k'}). \nn 
\eea
Taking the average over disorder the term with $j\neq j'$ disappears\footnote{Substituting the mean for the typical value of this sum is justified if the second sum over $j,j'$ is small, and this is not correct in $d=1,2$, although impossible to see in this simple calculation.} and we are left with
\bea
\Gamma_k &=&\pi \sum_{k'}\frac{W^2}{12L^d}\delta(E_k-E_{k'}) \nn \\ 
&=&\pi\int\frac{d^dk'}{(2\pi)^d}\frac{W^2}{12}\delta(E_k-E_{k'}) \nn \\  
&=&\pi\int dE' \rho(E')\frac{W^2}{12}\delta(E_k-E')=\frac{\pi W^2}{12}\rho(E_k). \nn 
\eea
The semiclassical velocity is that of the wave packet $v=\partial E_k/\partial k$. The formal expression for the diffusion coefficient is then $D=v^2\tau$, dropping explicit $k$ dependence in favor of the energy $E$
\bea
D=\frac{12\ v(E)^2}{\pi W^2\rho(E)}\sim \frac{g^3}{W^2}.
\eea
The diffusion coefficient (in units of lattice spacing) is inversely proportional to $W$ but never zero.

Remarkably\footnote{Let us mention that anomalous transport also exist in classical systems, in particular in low dimensions and in the presence of strong interactions between particles and/or if the disorder potential is unbounded (see e.g. Sinai's random walk in one dimension). However the nature of classical anomalous transport is not the same as Anderson localization. Anderson localization is indeed peculiar of the dynamics of a {\it single} particle in a {\it bounded} random potential.}, Anderson predicted in his seminal work \cite{Anderson1958} that the diffusion coefficient $D$ might in fact vanish for a sufficiently strong (but finite) disorder $W>W_c$, see Fig.~\ref{Fig:AndersonModel}. This is the phenomenon of Anderson localization, which thus states that the basin of attraction of the infinite disorder regime is finite: in the localized phase $\Delta x_t^2 $ and $|\psi(t,0)|^2$ remains finite for all $t$. 

Approaching the Anderson transition from the delocalized side the diffusion coefficient is seen vanishing and at the same time the local density of states becomes pure point, invalidating the use of Fermi golden rule. It is very difficult to do sufficiently accurate calculations in this way, usually one resorts to some kind of self-consistent equation in which the diffusion coefficient is found in a self-consistent way from the resummation of some diagrammatic series or equivalently the solution of a Dyson equation (see e.g. Chapter IV in \cite{akkermans2007mesoscopic}). This approach however is not controlled away from the small disorder limit and the results for the transition (e.g.\ critical exponents or multifractal properties of the eigenfunctions at the transition) can be deceptive. Perturbation theory in the disorder (locator expansion) (see Sec.~\ref{subsec:AL2:resolvent}) is better controlled, in particular in higher dimensions and for the many-body problem.

\subsection{Diagnostics of Anderson localization} \label{subsec:AL1:Diagno}

\paragraph{Dynamical characterization -- Absence of transport} The criterion for localization presented in the last section, i.e. a non-decay of the return probability for an initially localized excitation, is the one originally introduced by Anderson \cite{Anderson1958}. This is a characterization that gives to the Anderson localization transition the flavor of a dynamical phase transition. While it is not a static thermodynamic phase transition, it still has some influence on the macroscopic property of the system, in particular on transport properties. Indeed, through Einstein's relation, one expects the diffusion constant to be related to the conductivity $\sigma$  as $\sigma \sim D$. Another way to see why Anderson's Localization implies absence of transport is to use the Kubo formula for the conductivity. We will come back to this in the many-body case.

\paragraph{Eigenstate characterization -- Localized wave-functions}
The most striking characterization of the localized phase is perhaps that the eigenfunctions of the Hamiltonian $H$ resemble those of the infinite disorder case: they are localized around some localization center $j_{{\rm loc}} \in \JZ^d$, and are decaying exponentially away from it with a localization length $\zeta_{{\rm loc}}$: 
\bea
\psi_{{\rm loc}}(j) \sim e^{- |j-j_{{\rm loc}}|/ \zeta_{{\rm loc}}} \, .
\eea
Note that, similarly as in the infinite disorder case but in contrast with the free case, these eigenfunctions are true normalizable eigenfunctions of $H$ even in the $L \to \infty$ limit. A useful characterization of the localized character of an eigenfunction $\psi$ is to introduce its participation ratio
 \bea
I_q := \sum_{j} |\psi(j)|^{2q} \, .
 \eea
Taking $L$ finite, one expects the scaling $I_q \sim L^{d(1-q)}$ for a delocalized eigenfunction and $I_q \sim 1$ for a localized wave-function. This quantities are also useful to quantify more complex behavior: fractals wave-functions are also observed exactly at the transition between the localized and delocalized phase and a non-trivial fractal dimension $0<D(q)<d$ (multifractality) is introduced as $I_q \sim L^{D(q)(1-q)}$.

\paragraph{Spectral Characterization (1) -- Continuous Vs Point spectrum}
The spectrum of $H$, noted $\sigma(H)$ is by definition the set of energies $E \in \JR$ such that $H-E$ is not invertible. In this set, one should also distinguish: (i) the eigenvalues of $H$, corresponding to the `point-spectrum' $\sigma_p(H)$, which are the set of energies $E \in \JR$ such that there exist a vector $\ket{\psi} \in {\cal H}$ such that $H \ket{\psi} = E \ket{\psi}$; (ii) the rest of the spectrum that is called the {\it continuous spectrum}\footnote{One should also distinguish the absolutely continuous spectrum from the so-called singular spectrum, which are both parts of the continuous spectrum. We will not make that distinction here.} $\sigma_c(H)$. As seen in the previous example, this is a meaningful distinction in the thermodynamic limit $L \to\infty$. It also corresponds to a decomposition of the local density of state $\rho(j,E)$ previously introduced as, in general, $\rho(j,E) = \rho_c(j,E) + \rho_p(j,E)$. While $\rho_p$ (corresponding to the eigenvectors and to the point spectrum) is a discrete measure $\rho_p(j,E) := \sum_{\alpha \in {\cal I}} \delta(E-E_\alpha) |\braket{\alpha|j}|^2  $ with $E_\alpha$ (resp.\ $\ket{\alpha}$) the eigenvalues (resp.\ eigenvectors) of $H$ and ${\cal I}$ a countable set, $\rho_c$ is continuous with respect to the Lebesgue measure. To this decomposition also corresponds a decomposition of the Hilbert space as ${\cal H} = {\cal H}_p \oplus {\cal H}_c$, where ${\cal H}_p$ is generated by the eigenvectors of $H$ and ${\cal H}_c$ is defined as its orthogonal space. 

Spectral localization is the statement that the spectrum of $H$ only consists of a pure point spectrum as in the infinite disorder case. This notion is linked to the dynamical characterization of Anderson localization since the RAGE (Ruelle, Amrein, Georcescu and Enss) theorem (see \cite{HunzikerSigal2000} for review) ensures that states in ${\cal H}_c$ (resp. ${\cal H}_p$) correspond to conducting (resp. localized) states, see \cite{hundertmark2008} for details, and the precedent section for an elementary example in the infinite and zero disorder cases. Let us again insist that these notions only make sense in the thermodynamic limit $L\to \infty$. On a finite system (finite dimensional Hilbert space), $H$ is always diagonalizable and its spectrum is a discrete set. The discrete spectrum of a delocalized system converges in the $L \to \infty$ limit to an absolutely continuous spectrum, while it remains pure point for a localized system. In the following we will often not make the distinction between eigenvalues and elements of the spectrum, nor between eigenstate or `formal eigenstates' (e.g. plane waves) but this should not cause confusion.

\paragraph{Spectral characterization (2) -- Spectral statistics}

Anderson localization can also be seen directly from the study of the correlations between the eigenvalues of the Hamiltonian. In the localized phase, as in the infinite disorder limit $W \to \infty$, different eigenvalues are almost independent from one another (up to exponentially small corrections of the order of the overlap between eigenfunction). Arranging the energies $E_i$ in increasing order, the statistics of the normalized gap $s_i = \frac{E_{i+1}-E_i}{\langle E_{i+1} - E_i \rangle}$ converges (as $L \to \infty$) to a Poissonian\footnote{As an example take independent energy levels $\epsilon_1 , \cdots ,\epsilon_{N}$ ($N=L^d$) drawn from some distribution $p(\epsilon)$. Look at a given point $\epsilon_0$ that is included in the support of the distribution and consider $y:={\rm min} \{ \epsilon_i -\epsilon_0 , \epsilon_i-\epsilon_0>0\}$. In the large $N$ limit we rescale $y$ as $r:= N y$. The pdf of $r$ is evaluated as $Proba(r > s) = \left( 1- \int_{\epsilon_0}^{\epsilon_0 + s/N} p(\epsilon)d\epsilon \right)^N \sim_{N \to \infty} e^{-p(\epsilon_0) s} $. Hence, noting $\lambda := p(\epsilon_0)$ the pdf of $r$ is $p(r):=\lambda e^{-\lambda r}$. This obviously only works if we are at a point where $p(\epsilon_0) \neq 0$ (i.e. we should not look at the edge of the spectrum). Note also that the scaling factor $\lambda$ can be removed by a rescaling of $r$, but depends on the energy level that is being investigated.} statistics with $s_i$ distributed with the probability density function (pdf) $p(s) = e^{-s}$. In particular $p(0)$ is finite: no level repulsion. The presence of several eigenstates $\ket{\psi_1} , \ket{\psi_2}$ whose energies $E_1,E_2$ are arbitrarily close to one another is only possible in the localized phase where these eigenstates can also be far from one another in real space. In the delocalized phase eigenstates that would naively be arbitrarily close in energy are in fact interacting, and as a result hybridized by the hopping term of the Hamiltonian. We will see later on an elementary example of this hybridization but the important consequence for the energy statistics is that this leads to an effective repulsion between eigenvalues: $p(r=0)=0$. It is in fact believed (and supported by numerical evidence) that $p(r)$ is given in the $L \to \infty$ limit by the local level spacing statistics of a random matrix in the Gaussian orthogonal ensemble\footnote{The relevant random matrix ensemble in general depends on the symmetry of the problem.} (GOE): the so-called Wigner-Dyson distribution, and in particular $p(r)\sim_{r\to 0} r^2$.

\subsection{What is known} \label{subsec:AL1:Known}

Here we sate some known results on Anderson localization. These apply to the Anderson model as defined previously.

\paragraph{Localization at strong disorder in any $d$} It is known rigorously (see \cite{hundertmark2008} for an introduction to the rigorous aspects of the problem) that localization occur for strong enough disorder for $W>W_c$ with $W_c < \infty$ in any $d$. It is worth noticing that, although localization has been proven for sufficiently strong disorder, the absence of it has not been proven even in cases where we are sure, on numerical experiments grounds, that it does not exist for small disorder.

\paragraph{Localization in $d=1,2$} It is also known rigorously that in $d=1$, localization occurs for arbitrarily small disorder: $W_c(d=1)=0$. This can be seen (after Thouless \cite{Thouless1972}) using a transfer matrix method to solve the Schr\"odinger equation for the eigenvectors 
\bea
\begin{pmatrix}
\psi_{i+1}\\
\psi_i\end{pmatrix}=\begin{pmatrix}
\frac{(\epsilon_i-E)}{g} & -1\\
1 & 0\end{pmatrix}
\begin{pmatrix}
\psi_{i}\\
\psi_{i-1}\end{pmatrix},
\eea
or more formally
\begin{equation}
\Psi_{i+1}=T_i(E)\Psi_i.
\end{equation}

Proceeding on this route, one can eventually connect the localization length (which is related to the Lyapunov exponent of random products of the $T_i$'s) to the density of states in the celebrated Herbert-Jones-Thouless formula \cite{Thouless1972,Herbert-Jones1971}.

The same result is actually expected in $d=2$, $W_c(d=2) =0$. Although a rigorous solution of this problem does not exist, this claim was based first on the RG analysis of the {\it gang of four} \cite{AndersonThoulessAbrahamsFisher1980} and later confirmed by numerical investigations. The RG analysis is quite powerful and predicts that the Hamiltonian with a magnetic field (hermitian but not symmetric) does instead have a transition in $d=2$. This case is very much relevant for the theory of the Quantum Hall effect (see also \cite{ChalkerCoddington1988}).

\paragraph{Localization and delocalization in $d \geq 3$} The phenomenology becomes richer for $d\geq 3$. There there is a true phase transition at a finite value of the disorder strength $W_c >0$. This critical value furthermore depends on the energy $E$ of the eigenstates: parts of the spectrum might be localized (usually the tail of the spectrum, the `Lifshitz tails'), while other parts might remain delocalized. The critical disorder thus depends on $E$: $W=W_c(E)$. This defines `mobility edges'. The fact that the existence of both localized and delocalized eigenfunction at a given energy is forbidden is known as Mott's argument\footnote{The argument works as follows: consider two localized states close in energy, therefore far in space. If a delocalized eigenstate exists with energy between theirs then by varying the disorder realization very little and doing perturbation theory the two localized states will mix thanks to the delocalized state and therefore they will cease to be localized.}.

In $d\geq 3$ the phase transition is characterized by a diverging length scale: the localization length $\zeta_c(E)$ diverges as $W \to W_c(E)^+$. This suggests the existence of some universal properties and opens the way to some field theory approach to the phase transition ($(2+\epsilon)$ expansion), see \cite{EversMirlinReview} for review.

\section{Two analytical approaches to Anderson Localization}  \label{sec:AL2}

\subsection{The locator expansion for the resolvent} \label{subsec:AL2:resolvent}

In what follows we will be interested in showing localization in the Anderson model. As already mentioned, at a strictly infinite disorder strength $W= \infty$ localization is trivial: with probability $1$ the one-particle eigenvectors of $H$ are localized at some site $a \in \JZ^d$ with wave function $\psi_{a}(i) = \delta_{i,a}$ and energy $E = \epsilon_j$. The non-trivial statement is that this phase is stable. For $g/W \ll 1$ one would like to show that there is still an eigenfunction $\psi_{\alpha}$ roughly localized around $a$ and with energy $E_\alpha$: $\psi_{\alpha} =  \psi_a + o(1)$ and $E_{\alpha} = E_a + o (1)$. The most natural idea to show the existence of the localized phase is thus to expand around the trivial infinite disorder case. We thus formulate a perturbative expansion of the eigenfunction in the hopping term of the Hamiltonian (`locator expansion'). We split
\bea \label{Eq:SplittingAndersonHamiltonian}
&& H = H_0 + g T  \quad , \quad  H_0 :=\sum_{i }  \epsilon_i \ket{i} \bra{i} \quad , \quad  T := -  \left( \sum_{ \langle i j \rangle} \ket{i} \bra{j}  + \ket{j} \bra{i} \right) \, .
\eea
As is usual (and convenient) in this context we will start by formulating the perturbative expansion by considering the resolvent $G(E)$, defined as, $ \forall E \notin \sigma(H)$,
\bea
G(E) := \frac{1}{E-H} \,  .
\eea
We also note its matrix elements in the localized basis as $G(b,a,E) := \braket{b|G(E)|a}$. In the next section we will relate the resolvent to some useful physical observables, in particular to the potentially localized eigenfunctions. Considerations similar as to what follows can be e.g. found in \cite{hundertmark2008,PietracaprinaRosScardicchio2016,DeLucaPHD}.

\begin{figure}
\centerline{\includegraphics[width=12cm]{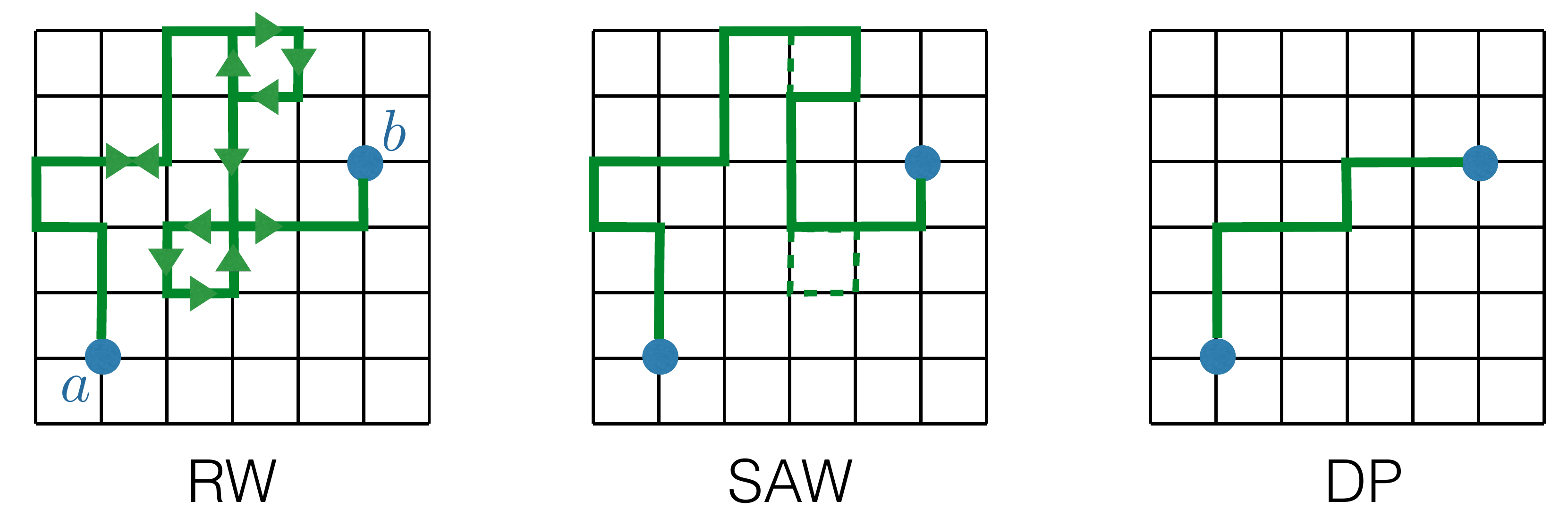}} 
\caption{The naive perturbation theory \eqref{Eq:Resolvent:RandomWalk} for the resolvent between $a$ and $b$ consists in summing all contributions from paths (random walks, left) between the two points. It always diverges due to local resonances. In the renormalized perturbation theory \eqref{Eq:Resolvent:SAWalk}, the sum is only on self-avoiding walks (middle), with the contribution of loops now resummed into self-energy corrections renormalizing the disorder. It diverges at the Anderson transition. In the forward approximation \eqref{Eq:WaveFunction:FWA}, the sum is restricted to the shortest paths (directed polymers, right), without taking into account the self-energies.}
\label{Fig:RW}
\end{figure}

\paragraph{Naive perturbation theory - Random-walk representation} We first obtain a representation of the resolvent in terms of an infinite series in the hopping $g$. To this aim we consider the unperturbed resolvent $G_0 (b,a,E)= \braket{b|\frac{1}{E-H_0}|a} = \delta_{ab} \frac{1}{E - \epsilon_a}$. Using the formula $A^{-1} - B^{-1} = B^{-1} (B-A) A^{-1}$ with $A = E - H$ and $B = E-H_0$ we obtain $G-G_0 = g G_0 T G$, i.e. $G = G_0 + g G_0 T G$ which, through iteration, leads to
\bea
G = G_0 +  \sum_{n=1}^{+ \infty}  g^n (G_0 T)^n G_0 \, . \nn 
\eea
Noting that $G_0$ is diagonal in the localized basis $\ket{a}$, $a \in \JZ^d$, and that $T$ only connects nearest neighbours, we obtain a {\it random-walk (RW) representation} for $G(a,b,E)$ as 
\bea \label{Eq:Resolvent:RandomWalk}
G(b,a,E) =  \frac{\delta_{a,b}}{E-\epsilon_a}  + \frac{1}{E-\epsilon_a}  \sum_{n=1}^{+\infty} \sum_{\pi : a \to b , |\pi| = n} (-g)^{n} \prod_{s=1}^n \frac{1}{E-\epsilon_{\pi(s)}}  \, ,
\eea
where here $\sum_{\pi : a \to b , |\pi| = n} $ is the sum over paths on the lattice of length $|\pi| = n  \geq 1$ with $\pi(0) = a$ and $\pi(n) = b$ (see Fig.~\ref{Fig:RW}).

\paragraph{Renormalized perturbation theory - Self-avoiding-walk representation} Consider now $b \neq a$ and the series (\ref{Eq:Resolvent:RandomWalk}). Here we want to transform the sum over random-walks into a sum over {\it self-avoiding walks} (see Fig.~\ref{Fig:RW}). The motivation for this will become clear in the next section. We thus need to factorize in this series all the loops. This can be done recursively: we first factorize all the walks from $a$ to $b$ which passes by $a$. This leads to
\bea
G(b,a,E) = G(a,a,E) \sum_{\pi : a \to b , |\pi| = n , \pi(s) \neq a} (-g)^{n} \prod_{s=1}^n \frac{1}{E-\epsilon_{\pi(s)}} \, , \nn 
\eea
where now the sum over paths is restricted to paths that do not come back to $a$. We can keep iterating this and we obtain
\bea
G(b,a,E) = G(a,a,E) \sum_{{\rm SAW} \pi : a \to b , |\pi| = n}  (-g)^{n} \prod_{s=1}^n   G(\pi(s) , \pi(s)  , E  )^{\JZ^d - \{\pi(0) , \cdots , \pi(s-1) \}} \, , \nn 
\eea
where $G(\pi(s) , \pi(s)  , E  )^{\JZ^d - \{\pi(0) , \cdots , \pi(s-1) \}}$ denotes the resolvent between $\pi(s)$ and $\pi(s)$ in the system where the point $\{\pi(0) , \cdots , \pi(s-1) \}$ have been removed. Removing the points recursively is necessary since paths that passes by $\pi(s')$ for $s' < \pi(s)$ should not be counted again since they have already been counted in $G(\pi(s') , \pi(s')  , E  )^{\JZ^d - \{\pi(0) , \cdots , \pi(s'-1) \}}$. This is usually reformulated by introducing the {\it self-energies} $\Sigma$ defined by
\bea
G(\pi(s) , \pi(s)  , E  )^{\JZ^d - \{\pi(0) , \cdots , \pi(s-1) \}} =: \frac{1}{E - \epsilon_{\pi(s)} - \Sigma_{\pi(s)}^{\{\pi(0) , \cdots , \pi(s-1) \}}(E)}  \, , 
\eea 
the self-avoiding walk (SAW) representation is thus rewritten as
\bea \label{Eq:Resolvent:SAWalk}
G(b,a,E) =  \sum_{{\rm SAW} \, \pi : a \to b , \,  |\pi| = n}  (-g)^{n} \prod_{s=1}^n   \frac{1}{E - \epsilon_{\pi(s)} - \Sigma_{\pi(s)}^{\{\pi(0) , \cdots , \pi(s-1) \}}(E)}   \, .
\eea
Note that the self-energies converge to $0$ in the infinite disorder limit: $\Sigma_{\pi(s)}^{\{\pi(0) , \cdots , \pi(s-1) \}}(E) \to_{g/W \to 0} 0$ (with probability $1$).

\subsection{Resonances and Hybridization - Convergence and divergence of the expansions} \label{subsec:AL2:resonances}

The RW expansion of the resolvent (\ref{Eq:Resolvent:RandomWalk}) is always (i.e. with probability $1$ in the thermodynamic limit) divergent. The reason for this is the existence of local {\it resonances}. 
A two-site resonance between two neighbouring sites $i$ and $j$ is said to occur if  $\frac{g}{\epsilon_i - \epsilon_j} \geq 1$. While this quotient is typically small and of order $g/W$, there is a probability of order $g/W$ that it is larger than $1$ (in fact the mean value of this quotient is infinite), which makes their existence inevitable in the thermodynamic limit. The existence of such a resonance implies that the terms in (\ref{Eq:Resolvent:SAWalk}) corresponding to paths passing from $i$ to $j$ (and $j$ to $i$) $m$ times diverge exponentially with $m$ since they contain a factor $(\frac{g}{\epsilon_i - \epsilon_j})^m$. 

\medskip

Physically this divergence is however not dangerous for localization. It just signals that any wave-function having a non-zero weight on one of the two sites will necessary have a weight on the other one (local delocalization). Locally the Hamiltonian looks like (restricted to the two-dimensional space ${\rm Vect}(\ket{i} , \ket{j})$)  the matrix 

\bea
H|_{(\ket{i} , \ket{j})} = \left( \begin{array}{cc}
	\epsilon_i & g  \\ 
	g & \epsilon_j  \end{array} \right) \, .
	\eea
	 Its exact eigenvalues are easily calculated as $\epsilon_{\pm} = \bar \epsilon \pm \sqrt{\delta \epsilon^2 + g^2 }$ with $\bar \epsilon = (\epsilon_i+\epsilon_j)/2$ and $\delta \epsilon = (\epsilon_j -\epsilon_i)/2$. The perturbative expansion in $g$ of the eigenvalues (and eigenvectors) of this Hamiltonian only converges for $|g| \leq |\delta|$. For $|g| \gg |\delta|$ one can think of locally replacing the basis in which one expects `localization' from $(\ket{i} , \ket{j}) \to (\ket{i}+ \ket{j} , \ket{i}-\ket{j})/\sqrt{2}$ and the random energies as $(\epsilon_1 , \epsilon_2 ) \to (\epsilon_+ , \epsilon_-)$. One says that the two states $(\ket{i} , \ket{j})$ are {\it hybridized} by the coupling. Note that then by construction $\epsilon_+ - \epsilon_- \geq 2 g$. This is the origin of level repulsion in the delocalized phase.

	 \medskip

The SAW representation does not suffer from these `trivial' local resonances, which have been taken care of through the introduction of the self-energies. Note that the SAW expansion is not `just' a perturbative expansion in $g/W$ since the self-energies depend on $g$, and in fact depend non-analytically on $g$ around $0$ due to the local resonances\footnote{There are poles arbitrarily close to 0 in the $g$ plane but their residue (the square of the amplitude of the wave function at $a$) is smaller the closer we get to zero, since the state responsible for the pole is at a distance diverging with $g\to 0$.}. The self-energies already take into account some non-perturbative features of the theory. However, the SAW expansion might still diverge (because of non-trivial resonances occurring at any scales), and its possible divergence is in fact considered since the work of Anderson \cite{Anderson1958} to be synonymous of delocalization. A convergent SAW is then synonymous of localization. The SAW expansion is still an expansion around the infinite disorder case, which is a good starting point in the localized phase.

\subsection{Some properties of the resolvent} \label{subsec:AL2:resolvent2}

\paragraph{Link between the resolvent and eigenfunctions} Taking first $L$ finite we expand the resolvent onto the eigenfunctions of $H$: denoting $\ket{\psi_{\alpha}}$/$E_{\alpha}$ the eigenvectors/energies of $H$, $G$ is rewritten as
\bea
G(b,a,E) = \sum_{\alpha} \frac{\psi_\alpha(b)\psi_\alpha^*(a)}{E-E_{\alpha}} \, . \nn
\eea
Hence, assuming no degeneracies, we obtain $\lim_{E \to E_{\alpha}} (E-E_{\alpha})G(b,a,E) = \psi_\alpha(b)\psi_\alpha^*(a) $. Combined with the SAW representation this leads to
\bea \label{Eq:SAW:WaveFunction}
\psi_{\alpha}(b) = \psi_{\alpha}(a) \sum_{{\rm SAW} \, \pi : a \to b , \,  |\pi| = n}  (-g)^{n} \prod_{s=1}^n   \frac{1}{E_{\alpha} - \epsilon_{\pi(s)} - \Sigma_{\pi(s)}^{\{\pi(0) , \cdots , \pi(s-1) \}}(E_\alpha)} \, .
\eea
This thus determines the amplitude of the wave-function at any point $b$ in terms of its value at another point $a$. In the localized phase one expects to be able to take the $L \to \infty$ of this, keeping $\psi_{\alpha}(a)$ (and thus $\psi_{\alpha}(b)$) of order $O(1)$. The convergence of the expansion then indeed implies localization.

\paragraph{Link between the resolvent and the continuous spectrum} Starting again with $L$ finite, we may also write $G(a,a,E) = \sum_{\alpha}   \frac{1}{E-E_{\alpha} }|\braket{\alpha|a}|^2$. In the large $L$ limit and in the presence of both continuous and point spectrum this converges to 
\bea
G(a,a,E) =_{L \to \infty} \int_{\omega \in \sigma_c(H)} d\omega  \frac{1}{E-\omega}\rho_c(a,\omega)  + \sum_{\alpha \in {\cal I}}   \frac{1}{E-E_{\alpha} }|\braket{\alpha|a}|^2 \, , \nn 
\eea 
where we have separated the contribution from the (localized) eigenvectors of $H$, $\ket{\alpha}$, from the sum onto the (delocalized) continuous part of the spectrum, reintroducing the local continuous part of the density of state $\rho_c(a,\omega)$. For $\eta>0$ small we compute:
\bea
&& G(a,a,E-i \eta)-G(a,a,E+i \eta) = \int_{\omega \in \sigma_c(H) + i \eta} d\omega  \frac{1}{E-\omega} \rho_c(a,\omega - i \eta)  \nn \\
&&  -  \int_{\omega \in \sigma_c(H) - i \eta} d\omega  \frac{1}{E-\omega} \rho_c(a,\omega + i \eta ) + \sum_{\alpha \in {\cal I}}  \frac{2 i \eta}{(E-E_{\alpha})^2 + \eta^2} |\braket{\alpha|a}|^2   \nn 
\eea
Now, note that for $E \neq E_{\alpha}$, the discrete sum is of order $\eta$. While it is indeed possible that $E=E_{\alpha}$ for some $\alpha$ (in which case the discrete sum is large, of order $1/\eta$), this can only occurs with probability $0$ if $E$ is chosen randomly inside the spectrum of $H$. Indeed the sum $\sum_{\alpha \in {\cal I}}$ is a sum over a countable set, and the set $\{E_\alpha\}$ is necessarily a set of measure $0$ (with respect to the Lebesgue measure) in $\sigma(H)$. Here typically ${\cal I}$ can be thought of as labeling the localization center and thus ${\cal I} \subset \JZ^d $. On the other hand the contribution of the continuous part of the spectrum can be rewritten as a contour integral:
\bea
&& \int_{\omega \in \sigma_c(H) + i \eta} d\omega  \frac{1}{E-\omega} \rho_c(a,\omega - i \eta)  -  \int_{\omega \in \sigma_c(H) - i \eta} d\omega  \frac{1}{E-\omega} \rho_c(a,\omega + i \eta ) \nn \\
&&  = \int_{\cal C}  \frac{1}{E-\omega} \left( \rho_c(a,\omega + i \eta ) +  \rho_c(a,\omega - i \eta )   \right) + O(\eta) \, , \nn
\eea
with ${\cal C}$ a positively oriented, rectangular contour of width $\eta$ around $\sigma_c(H)$. Hence, in the limit $\eta$ to $0$, we get, with probability $1$
\bea
\rho_c(a,E) = \frac{1}{2 i\pi} \lim_{\eta \to 0^+} \left(G(a,a,E - i \eta) -G(a,a,E + i \eta) \right) \, ,
\eea
or also  $\rho_c(a,E) = - \frac{1}{\pi}\lim_{\eta \to 0^+} \Im G(a,a,E + i \eta)$. The presence of continuous spectrum manifests itself in the existence of a branchcut in the resolvent. For example in the absence of disorder for the Anderson model in $d=1$ one has $G(0,0,E) = \int_{-\pi}^{\pi}  \frac{dk}{2 \pi} \frac{1}{2 + 2 g \cos(k)} = \frac{1}{\sqrt{E+2g} \sqrt{E-2g}}$. This indeed has a branchcut and $\rho_c(0,E) = \frac{1}{2\pi  \sqrt{4g^2 - E^2}} \theta(2 g- |E|)=- \frac{1}{\pi}\lim_{\eta \to 0^+} \Im G(0,0,E + i \eta)$ as found in Sec.~\ref{subsubsec:AL1:Intro:DisorderFree}. The absence of continuous spectrum, i.e. localization at the energy $E$, therefore implies that with probability $1$ $\lim_{\eta \to 0^+} \Im G(a,a,E + i \eta) = 0$.

\paragraph{Link between the resolvent and unitary evolution} The resolvent is related to the evolution operator $U(t):= e^{-i t H}$ as $U(t) = \int_{\cal C} \frac{dz}{2 i \pi} e^{-i t z} G(z)$ with ${\cal C}$ a positively oriented contour around $\sigma(H)$, or also as $\theta(t) U(t) = \lim_{\eta \to 0^+} - \int_{\JR} \frac{dz}{2 i \pi} e^{-it z} G(z + i \eta)$. In particular, starting from an initially localized function around some site $a$, $\psi(t=0,a)$, the return probability is computed as $p(t) = \lim_{\eta \to 0^+} - \int_{\JR} \frac{dz}{2 i \pi} e^{-it z} G(0,0,z+i\eta) $, i.e. $p(t) = \lim_{\eta \to 0^+} - \int_{\JR} \frac{dz}{2 i \pi} e^{-it z} (z+i\eta - \epsilon_a - \Sigma_a(z+i\eta))^{-1}$. From this expression it is perhaps not so difficult to convince oneself that the presence of continuous spectrum, i.e. $\lim_{\eta \to 0^+} \Im G(a,a,E + i \eta) \neq 0$ and thus $\lim_{\eta \to 0^+} \Im \Sigma_a(E + i \eta) \neq 0$ (branch cut), implies a decay of the return probability $p(t)$.

\subsection{Anderson localization on the Bethe lattice: self-consistent approach} \label{subsec:AL2:selfconsistent}

\begin{figure}
\centerline{\includegraphics[width=6cm]{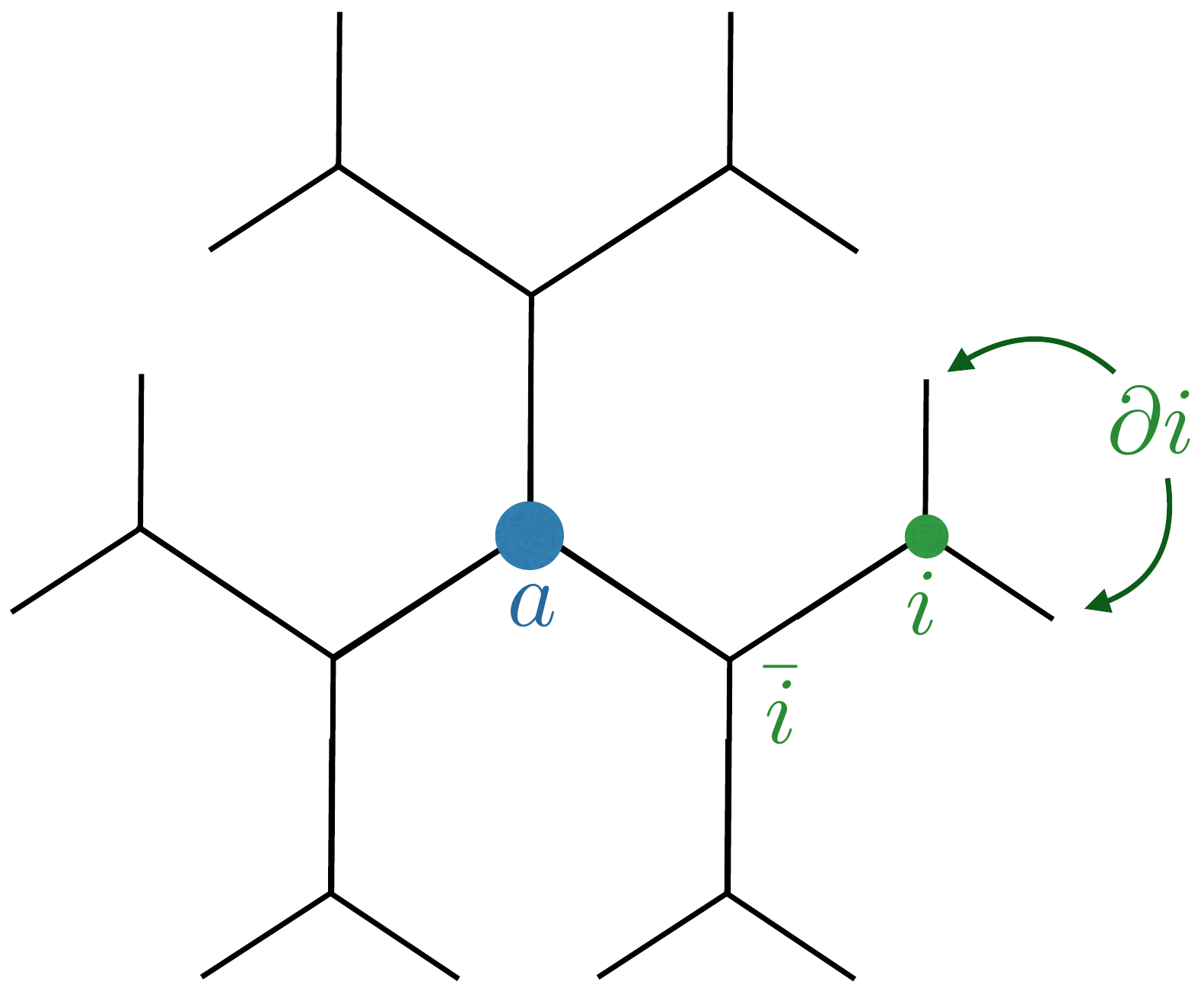}} 
\caption{Portion of the Bethe lattice with branching number $K=2$.}
\label{Fig:BetheLatt}
\end{figure}

We now turn to the analysis of Anderson localization on the Bethe lattice (infinite tree, i.e. planar graph without loops, where all vertices have a constant coordination number $z = K+1$, see Fig.~\ref{Fig:BetheLatt}) using the self-consistent approximation originally introduced in \cite{AbouChacraThoulessAnderson1973a}. This approach is essentially identical to the approach used by Anderson in \cite{Anderson1958} to study localization on $\JZ^d$ (second order perturbation theory for the self-energies), but the approximation used there are better justified on the Bethe lattice, a case on which we thus focus.

\paragraph{Derivation of the self-consistent approximation}
 Let us start by remarking that the SAW representation of the resolvent (\ref{Eq:Resolvent:SAWalk}) simplifies in the case of the Bethe lattice. Indeed, since there is no loop, the sequence of self-energies $\Sigma_{\pi(s)}^{\{\pi(0) , \cdots , \pi(s-1)\}}$ is `simply' equal to $\Sigma_{\pi(s)}^{\{\pi(s-1)\}}$: once the edge connecting $\pi(s)$ to $\pi(s-1)$ has been removed a path emanating from $\pi(s)$ cannot visit the points $\pi(u)$ for $u \leq s-1$. We can thus rewrite

\bea
G(b,a,E) =  \sum_{{\rm SAW} \,  \pi : a \to b , \, |\pi| = n}  (-g)^{n} \prod_{s=1}^n   \frac{1}{E - \epsilon_{\pi(s)} - \Sigma_{\pi(s)}^{ \{ \pi(s-1) \}}(E)}   \, . \nn 
\eea
At this stage this is exact. The idea is now to obtain the statistics of the self-energies $ \Sigma_{\pi(s)}^{ \{ \pi(s-1) \}}(E)$ in a self-consistent approximation. Let us first simplify the notations by adopting a sense of reading on the Bethe lattice, i.e. decide that the lattice is rooted at the point $a$, and note $\Sigma_{i}^{\circ}(E)$ the self-energy at site $i$ on the Bethe lattice with the edge going from site $i$ in the direction of the root removed (see Fig.~\ref{Fig:BetheLatt}). We also denote $\partial i$ the descendants of $i$ and $\bar i$ the father of $i$ according to this sense of reading. Using again the RW representation of the resolvent (\ref{Eq:Resolvent:RandomWalk}) we have by definition of $\Sigma_{i}^{\circ}(E)$
\bea
E-\epsilon_i - \Sigma_{i}^{\circ}(E) = \left[\frac{1}{E-\epsilon_i} \left( 1 + \sum_{n=2}^{\infty}  \, \, \sum_{{\rm RW} \,  \pi : i \to i, \, |\pi| = n , \,  \pi(u) \neq \bar i } (-g)^n \prod_{u=1}^n \frac{1}{E-  \epsilon_{\pi(u)}}  \right)  \right]^{-1} \nn 
\eea 
Now this sum can formally be restricted to paths that only visit the sites $i$ and its descendants $\partial i$ by reintroducing some self-energies. These paths are of even length and 
\bea
&& E-\epsilon_i - \Sigma_{i}^{\circ}(E) = \nn  \\
&& \!\!\!\!\!\!\!\!\!\!\!\!\!  \left[\frac{1}{E-\epsilon_i} \left( 1 + \sum_{m=1}^{\infty} \, \, \sum_{{\rm RW} \pi : i \to i, \,  |\pi| = 2m , \, \pi(u) \in \{ i\} \cup \partial i } (-g)^n \prod_{u=0}^m \frac{1}{E-  \epsilon_{\pi(2u+1)} - \Sigma_{\pi(2u+1)}^{\circ}(E)} \frac{1}{E-\epsilon_i}  \right)  \right]^{-1}  \, , \nn 
\eea 
which is exact if the sum converges. Now the approximation consists in restricting this sum to paths of length $2$, i.e. paths with $\pi(0) = \pi(2) = i$ and $\pi(1) \in \partial i$. There are exactly $K$ such paths and expanding in $g$ then leads to
\bea
E-\epsilon_i - \Sigma_{i}^{\circ}(E) = E-\epsilon_i \left( 1 - g^2 \sum_{j \in \partial i} \frac{1}{E-\epsilon_j - \Sigma_{j}^{\circ}(E)} \frac{1}{E-\epsilon_i} \right)  \, , \nn 
\eea
i.e.
\bea \label{Eq:SCA:SelfE}
 \Sigma_{i}^{\circ}(E)  = g^2  \sum_{j \in \partial i} \frac{1}{E-\epsilon_j - \Sigma_{j}^{\circ}(E)} \, .
\eea

\paragraph{Localization threshold in the self-consistent approximation}
Eq.~(\ref{Eq:SCA:SelfE}) can be turned to a self-consistent equation for the pdf of the self-energies $\Sigma_{i}^{\circ}(E)$ by using the fact that the self-energies $\Sigma_{i}^{\circ}(E)$ on the lhs and $\Sigma_{j}^{\circ}(E)$ on the rhs should be identically distributed (note that the different $\Sigma_{j}^{\circ}(E)$ appearing on the rhs of (\ref{Eq:SCA:SelfE}) are independent). This equation can be found in \cite{AbouChacraThoulessAnderson1973a} but cannot be solved analytically. The existence of a localized phase can however be easily investigated using some approximations. Noting $E_i$ and $ \Delta_i$ the real and imaginary parts of the self-energy: $ \Sigma_{i}^{\circ}(R+i\eta) = E_i(R+i\eta) - i \Delta_i(R+i\eta)$, we check the conditions for the imaginary part $\Delta_i$ to be small (almost surely of order $\eta$) when $\eta$ is going to $0$. The existence of such a distribution is taken as a necessary condition for localization. The reason why this should be interpreted as localization is because, as we saw in Sec.~\ref{subsec:AL2:resolvent2}, the absence of continuous spectrum at the energy $R$ implies that $\lim_{\eta \to 0} \Im G(a,a, R+ i \eta) =0$ (almost surely). Because of the link between the resolvent and the self-energy, this also implies that $\lim_{\eta \to 0} \Im \Sigma_a(R+ i \eta) = 0$ (almost surely) in the localized phase.

The self-consistent equation for the real and imaginary part of the self-energies on the Bethe lattice reads, probing the spectrum around the energy $R$: $E=R+i\eta$
\bea
&& E_i = g^2 \sum_{j=1}^k \frac{R-\epsilon_j-E_i}{(R-\epsilon_j-E_j)^2 + (\eta+\Delta_j)^2} \sim g^2 \sum_{j=1}^k \frac{1}{R-\epsilon_j -E_j}  \nn \\
&& \Delta_i = g^2 \sum_{j=1}^k \frac{\eta+ \Delta_j}{(R-\epsilon_j-E_j)^2 + (\eta+\Delta_j)^2} \sim  g^2 \sum_{j=1}^k \frac{\eta+ \Delta_j}{(R-\epsilon_j-E_j)^2} \, , \nn 
\eea
where on the right-hand side we have assumed that we are in the localized phase and that $\Delta$ is small, of order $\eta$. Neglecting also the $E_j \sim g^2/W$ in the denominator in the expression for $\Delta_j$ implies that $E_i$ and $\Delta_i$ can be considered as independent. The distribution of $\Delta_i$ then satisfies the equation (changing $\Delta_i \to \eta \Delta_i$)
\be
P(\Delta_0) = \int_{\epsilon_i  \in [-W/2,W/2]^d} \frac{d\epsilon_1 \cdots d \epsilon_K}{W^K}   \int d\Delta_1 \cdots d\Delta_K P(\Delta_1) \cdots P(\Delta_K) \delta\left(\Delta_0 - g^2  \sum_{j=1}^k \frac{1+ \Delta_j}{(R-\epsilon_j)^2} \right) \,, \nn 
\ee
and we investigate the condition on $W$ for such a distribution to exist. Taking the Laplace transform $f(s)= \int_{0}^{\infty} e^{-s \Delta} P(\Delta) d \Delta$ we obtain
\bea
f(s) =  \int_{-W/2}^{W/2} \frac{d\epsilon}{W} \, f\left(g^2 \frac{s}{(R-\epsilon)^2}\right) e^{-\frac{st^2}{(R-\epsilon)^2}}    \, . \nn 
\eea
Now, since $\Delta_i$ contains $1/(\epsilon-R)^2$, the pdf $P(\Delta)$ satisfies $P(\Delta) \geq A/|\Delta|^{3/2}$ for $|\Delta| \to \infty$ and $A$ some constant. Assuming thus $P(\Delta) \sim A |\Delta|^{-\alpha}$ with $1  \leq \alpha \leq 3/2 $ (normalizability) one obtains that at small $s$, $f(s) =1 - A' s^{\beta}$ with $\beta = \alpha-1 \in [0,1/2] $. Hence
\bea
1-A' s^{\beta} && =\left(1 - A' \int_{R-W/2}^{R+W/2} \frac{dx}{W}  \frac{g^{2\beta}}{x^{2\beta}} s^{\beta} + \cdots  \right)^K \nn \\
&& \simeq 1 -  A' s^{\beta} K g^{2\beta} \int_{R-W/2}^{R+W/2} \frac{dx}{W}  \frac{1}{x^{2\beta}} \nn 
\eea
And therefore we obtain
\bea
 K g^{2\beta} \int_{R-W/2}^{R+W/2} \frac{dx}{W}  \frac{1}{x^{2\beta}} =  1 \, . \nn 
\eea
In the center of the band $R=0$ this reads 
\bea
1= \frac{K}{1-2\beta} \left( \frac{2g}{W} \right)^{2\beta} \, . \nn
\eea
Noting $f(\beta,z) = \frac{K}{1-2\beta} (2 z)^{2\beta} $ we are interested in the existence of solutions of $f(\beta,z) = 1$ for $\beta \in [0,1/2]$ and $z \geq 0$. For $z$ close to $0$ (large disorder) the solution is close to $\beta =1/2$: writing $\beta = 1/2 - \delta \beta$ one obtains $\delta \beta \simeq \frac{Kz}{2} = \frac{Kg}{2W}$. The solution $z(\beta)$ exist for $\beta \leq \beta_c$ and $z(\beta)$ increases as $\beta$ decreases: $z'(\beta) = \partial_\beta f/\partial_z f|_{z=z(\beta)} \leq 0$. The solution ceases to exist when $\partial_\beta f|_{z=z(\beta)} = 0$. This occurs at $\beta=\beta_c$ that thus satisfies $\ln(2 z_c) = (1-2\beta_c)^{-1}$. Hence the critical disorder strength is given by \cite{AbouChacraThoulessAnderson1973a}
\bea
\frac{W_c}{g} = 2 K e \ln \left( \frac{W_c}{2g} \right) \, .
\eea
This should be compared with the exact result that reads $\frac{W_c}{g} = 4 K \ln \left( \frac{W_c}{2g} \right)$ (see \cite{tarquini2017critical}).

\subsection{Forward scattering approximation} \label{subsec:AL2:FWA}

We now investigate localization in the Anderson model using the so-called forward scattering approximation. Assuming that we are in the localized phase and considering a wave-function $\psi_{\alpha}$ that in the limit $g\to0$ converges to the wave-function localized at site $a$, $\lim_{g\to0}\psi_{\alpha}(j)=\delta_{j,a}$, we obtain from (\ref{Eq:SAW:WaveFunction}) that to lowest order in $g$ (using also $E_\alpha \to_{g \to 0} \epsilon_a$ and $\Sigma \to_{g \to 0} 0$),
\bea \label{Eq:WaveFunction:FWA}
\psi_{\alpha}(b) \simeq \sum_{{\rm DP}\, \pi:a\to b }  (-g)^{n} \prod_{s=1}^n   \frac{1}{\epsilon_{\alpha} - \epsilon_{\pi(s)}}   \,,
\eea
where now the sum $\sum_{{\rm DP}\, \pi:a\to b}$ is only over the {\it shortest paths from $a$ to $b$}, also called {\it Directed Polymers} (DP), and $n =|\pi|$ denotes the length of such a path; see Fig.~\ref{Fig:RW}. The approximation (\ref{Eq:WaveFunction:FWA}) of the wave-function is known as the {\it Forward Scattering approximation} (FWA). This approximation thus consists in approximating $\psi_{\alpha}(b)$ by its lowest order in $g$ (using naive perturbation theory).

\medskip

The passage from the exact SAW representation (\ref{Eq:SAW:WaveFunction}) to the FWA involves: (i) neglecting the contribution of longer paths; (ii) neglecting the self-energy correction. It is often argued, and confirmed by numerical simulations \cite{PietracaprinaRosScardicchio2016}, that the FWA overestimates the probability of presence $|\psi_{\alpha}(b)|^2$, although the estimation gets better as the dimension of the lattice increases \cite{PietracaprinaRosScardicchio2016}, suggesting that it is a mean-field theory for Anderson localization. Localization is thus harder in the FWA than it really is, and in general the critical disorder value causing localization in the FWA is larger than the true value: $W_c^{{\rm FWA}} > W_c$. In particular, while it is an exact result that $W_c = 0$ in $d=1$, we will see that there $W_c^{{\rm FWA}}>0$. One reason to expect that is that the introduction of self-energies `kill' the possible small energy denominators through exact local diagonalization of the Hamiltonian: taking care of the resonance (see discussion below Eq.~(\ref{Eq:Resolvent:SAWalk})) can be thought of as the replacement $\epsilon_i - \epsilon_j \to \epsilon_+ - \epsilon_- \geq 2 g$ (`avoided level crossing'). Still, the FWA does not diverge as easily as the RW (which always does) representation since passing by several times on a local resonance is not possible in the SAW.

\medskip

In the remaining of this section we work out explicitly the forward scattering approximation in two cases: in dimension $1$ and on the Bethe lattice\footnote{The general case is still difficult to tackle analytically, even in this approximation, since the sum of random variables in (\ref{Eq:WaveFunction:FWA}) is strongly correlated. On the other hand (\ref{Eq:WaveFunction:FWA}) is well suited to numerical studies, even in high dimensions, since (\ref{Eq:WaveFunction:FWA}) can be computed using a transfer-matrix algorithm that is $O(n^{d})$, see \cite{PietracaprinaRosScardicchio2016}. Let us mention here the connection between the FWA and the directed polymer (DP) problem (exactly solvable in $d=2$), that is the equilibrium statistical mechanics of directed paths in a random environment. Indeed the DP is formally equivalent to the FWA is one replaces $(\epsilon_{\alpha} - \epsilon_{\pi(s)})^{-1}$ in (\ref{Eq:WaveFunction:FWA}) with $e^{-E_{\pi(s)}/T}$ with $T>0$ and $E_{i}$ some iid random energies. The caveat here is of course that $(\epsilon_{\alpha} - \epsilon_{\pi(s)})^{-1}$ can be both positive and negative. Still, see \cite{MedinaKardar1992,SomozaOrtunoPrior2007,SomozaLeDoussalOrtuno2015,PietracaprinaRosScardicchio2016} for more on this interesting connection.}. As a criterion for localization we will here directly investigate the condition on $W$ for the `potentially localized' wave-function around the point $a$, $\psi_{\alpha}$, to indeed be localized: $|\psi_{\alpha}(b)|^2$ decays exponentially with the distance. The criterion\footnote{One might worry that this criterion is insufficient to ensure the convergence of the expansion on the Bethe lattice. We comment on that in the end of Sec.~\ref{subsubsec:AL2:FWABethe}.} is thus that there exist $\alpha>0$ such that
\bea \label{Eq:LocCriterion1}
\lim_{n \to \infty} \JP\left( {\rm max}_{b | {\rm dist}(a,b)=n} \frac{\log |\psi_{\alpha}(b)|^2}{n} <-\alpha \right) = 1 \,.
\eea
And the maximum value of $\alpha$ for which this holds defines the localization length 
\bea
\zeta = 1/\alpha_{\rm max} \, .
\eea
The latter can in principle depend on the chosen states, e.g. through its energy. Note that it is important to take the infinite volume limit first here, keeping $|\psi_{\alpha}(a)|^2$ of order $1$.

\subsubsection{Localization in $d=1$ through the FWA} \label{subsubsec:AL2:FWAd1}

In $d=1$ the index $i$ is now an integer $i \in \JZ$, and there is a single path of minimal length connecting the origin $0$ to any-point. The FWA approximation of the amplitude at point $n \in \JN$ of the wave-function that is potentially localized at $0$ is thus
\bea \label{Eq:FWAd1.1}
| \psi_0(n) |^2 = (g)^{2n} \prod_{i=1}^n   \frac{1}{(\epsilon_{0} - \epsilon_{i})^2}  \, . 
\eea
In what follows we take for simplicity $\epsilon_0 =0$, and therefore investigate localization `in the middle of the band'. In order to check the localization criterion (\ref{Eq:LocCriterion1}) we need to study the behavior of the random variable
\bea
\frac{1}{n} \ln | \psi_0(n) |^2 = \ln \left( 2g/W \right)^2  + \frac{1}{n}  \xi_n  \, , \nn 
\eea
where $\xi_n$ is a sum of iid RVs: $\xi_n:= \sum_{j=1}^{n} \ln y_j$ with the $y_j \geq 1 $ iid RVs distributed as $1/z^2$ with $z$ a RV uniformly distributed in $[-1,1]$. Hence the $y_j$ are distributed with the pdf $p(y) = \frac{1}{2 y^{3/2}} \theta(y-1)$. Without doing much calculations, note that the central limit theorem (CLT) trivially implies that $\xi_n$ is distributed as a Gaussian random variable\footnote{While the distribution of $y_j$ has a fat-tail, that is not the case for the distribution of $\ln y_j$.} in the large $n$ limit, centered around $n  \int_1^{+\infty} \log(y) p(y) dy = n \int_0^{+\infty}  (u/2) e^{-u/2} du = 2 n$, with fluctuations of order $\sqrt{n}$. This means that the localization length is deterministic and given by
\bea
\zeta^{-1} = -\lim_{n \to \infty} \frac{1}{n} \ln | \psi_0(n) |^2  = - \ln \left( 2g/W \right)^2 - 2 \quad , \quad \text{almost surely.} 
\eea
This localization length is finite for $W>W_c^{{\rm FWA}}$ with $W_c^{{\rm FWA}}/g = 2 e$. Note that, as said before, Thouless showed that $W_c=0$ in $d=1$: the FWA does indeed underestimate the extension of the localized phase. 

\smallskip

Before we continue let us note that the average of $| \psi_0(n) |^2 $ as given in (\ref{Eq:FWAd1.1}) is clearly infinite $\forall W$. On the other hand we typically have $| \psi_0(n) |^2 \sim e^{-n/\zeta}$ in the localized phase. This confusing fact is due to the fat tail in the distribution of $1/\epsilon_i$ which makes the average value of $| \psi_0(n) |^2 $ meaningless. The take-home message is that one has to be careful with average values since distributions with fat tails are ubiquitous in Anderson localization due to the presence of energy denominators.

\subsubsection{Localization on the Bethe lattice through FWA} \label{subsubsec:AL2:FWABethe}

We now consider the Anderson model on the Bethe lattice with coordination number $K\geq2$ ($K=1$ is just the already treated one-dimensional case). There are now ${\cal N}_n:=(K+1)K^{n-1}$ points at a distance $n$ of a given point and we need to check the localization criterion (\ref{Eq:LocCriterion1}): in the localized phase there exist $\alpha>0$ such that
\bea
\lim_{n \to \infty}  \JP\left( {\rm max}_{i \in\{ 1 , \cdots ,\, {\cal N}_n\}}  \frac{\xi_n^i - \Theta_n}{n} < -\alpha \right) = 1 \, ,  \nn 
\eea
where the $(\xi_n^i)_{i \in \{1 , \cdots , \, {\cal N}_n \}} $ are iid random variables distributed as $\xi_n$ before and $\Theta_n := n \theta$ with $\theta:= 2 \ln(W/(2g))$. In order to check this we now really need the distribution of $\xi_n$. Indeed, while it is still true that for a finite number $M$ fixed of $\xi_n^i$, the maximum ${\rm max}_{i\in \{1,\cdots ,\, M\}} \xi_n^i/n$ is deterministic and equal to $2$ in the large $n$ limit, here we are interested in the maximum of an exponential number in $n$ of iid $\xi_n^i$. Such a calculation probes the distribution of $\xi_n$ out of the range of application of the CLT and large deviations become important. Let us see that explicitly: (assuming below that we are in the localized phase, and thus that the computed probability is arbitrarily close to $1$ for $n$ large)
\bea
\JP\left( {\rm max}_{i \in\{ 1 , \cdots ,\, {\cal N}_n\}}  \frac{\xi_n^i - \Theta_n}{n} < -\alpha \right)  &=& \JP\left(\frac{\xi_n}{n} < -\alpha+\frac{\Theta_n}{n}\right)^{{\cal N}_n} \nn \\
&=&\left[1-  \JP\left(\frac{\xi_n}{n} > -\alpha+\frac{\Theta_n}{n})\right)\right]^{{\cal N}_n} \nn \\
&=& \exp\left[{\cal N}_n \ln\left( 1-  \JP\left(\frac{\xi_n}{n} > -\alpha+\frac{\Theta_n}{n}\right)  \right)  \right] \nn \\
&\simeq& \exp\left[ - {\cal N}_n \JP\left(\frac{\xi_n}{n} > -\alpha+\frac{\Theta_n}{n}\right)  \right] \nn
\eea
And this must converge to $1$. While it is true that $\JP\left(\frac{\xi_n}{n} > -\alpha+\frac{\Theta_n}{n}\right) \to_{n \to \infty} 1$ for $\alpha>\theta-2$ and $\JP\left(\frac{\xi_n}{n} > -\alpha+\frac{\Theta_n}{n}\right) \to_{n \to \infty} 0$ for $\alpha<\theta-2$, this is not sufficient to conclude that the localization length is $\zeta^{-1} = \theta-2$ and that critical disorder is obtained for $\theta=2$ (this is the result in $d=1$). Indeed, for $\alpha<\theta-2$ the probability only decays exponentially (this will be shown below)
\bea
\JP\left(\frac{\xi_n}{n} > -\alpha+\frac{\Theta_n}{n}\right)  \sim_{n \to \infty} e^{-n \phi(\alpha)} \, , \nn 
\eea
with $\phi(\alpha)$ a large deviation function with $\phi(\alpha) >0$ for $\alpha <\theta-2$ and $\phi(\alpha)=0$ for $\alpha=\theta-2$. Since ${\cal N}_n \sim K^n$ there can be a competition in the exponential and we need to check that $\lim_{n \to \infty} \exp(-e^{n (\log(K) - \phi(\alpha))}) = 1$. Hence we actually need to have $\phi(\alpha) > \log(K)$ for all $\alpha\in ]0,\alpha_c[$ in the localized phase with $0<\alpha_c<\theta-2$. To compute the large deviation function we first obtain directly the full pdf of $\xi_n$, noted $P_n(\xi)$. It can be computed using Laplace transform:
\bea
g_n(s) &:=& \int_{0}^{\infty} d \xi P_n(\xi) e^{-s \xi}  \nn \\
&=& \int_{0}^{\infty} d\xi \int_{y_1=0}^{\infty} dy_1 \cdots \int_{y_n=0}^{\infty} dy_n \delta(\xi - \sum_{i=1}^{n} \ln(y_i)) e^{-s \xi} p(y_1) \cdots p(y_n) \nn \\ 
&=& \left(  \int_1^{\infty} dy p(y) e^{-s \ln(y)}\right)^n \nn \\
&=& \left(  \frac{1}{1+2s} \right)^n  \, . \nn 
\eea 
Taking the inverse Laplace transform we thus obtain, 
\bea
&& P_n(\xi) = \int_{-\frac{1}{2}^- - i \infty}^{-\frac{1}{2}^- +i \infty} \frac{ds}{2 i \pi} e^{s \xi} \left(  \frac{1}{1+2s} \right)^n  \nn \\
&& = \int_{-\frac{1}{2}^- - i \infty}^{-\frac{1}{2}^- +i \infty} \frac{ds}{2 i \pi} e^{-\xi/2} \sum_{m=0}^{\infty} \xi^m \frac{(s+1/2)^m}{m!} \left(  \frac{1}{1+2s} \right)^n    \nn
\eea
Hence
\bea
&& P_n(\xi) =  \frac{\xi^{n-1}}{2^n(n-1)!} e^{-\xi/2} \theta(\xi) \, .
\eea
For $\alpha < \theta-2$ we thus have
\bea
\JP\left(\frac{\xi_n}{n} > -\alpha+\frac{\Theta_n}{n}\right) &=& \int_{-n\alpha + \Theta_n}^{\infty} P_n(\xi) d\xi  \nn \\
& \simeq & P_n(n(\theta-\alpha)) \nn \\
& \sim & \exp\left((n-1)\log(n(\theta-\alpha)) - n \log(2) - n \frac{\theta-\alpha}{2} -(n-1)\log((n-1)/e)  \right) \nn \\
& \sim & \exp(-n\phi(\alpha)) \nn
\eea
with the large deviation function $\phi(\alpha) = \frac{\theta-\alpha}{2} - \log\left(  \frac{\theta-\alpha}{2}  e \right)$ and we have only retained here the main (exponential) dependence on $n$. In the localized phase the localization length  $\zeta^{-1}$ is thus defined as the unique real positive solution of $\phi(\zeta^{-1}) = \log(K)$. Approaching the critical point $\zeta$ diverges and the position of the critical point is determined by $\phi(0)= \log(K)$, i.e
\bea \label{Eq:ResBetheLatticeWcFWA}
\frac{W_c^{{\rm FWA}}}{g} = 2 e K \ln\left(\frac{W_c^{{\rm FWA}}}{2g}\right)  \, .
\eea
This is the same result than the one obtained in the self-consistent approximation.

\paragraph{Remark on the localization criterion} On the Bethe lattice one might worry that the fact that the number of sites at a distance $n$ of a given site $a$ on the Bethe lattice grows exponentially might render the forward approximation of the wave-function non-normalizable even when the localization criterion (\ref{Eq:LocCriterion1}) is satisfied. Indeed, the convergence of the expansion of the wave-function in the FWA must imply that, with probability $1$,
\bea \label{Eq:CaveatBetheLattice}
\lim_{n^* \to \infty} \sum_{n \geq n^*} \,  \sum_{b , \,  {\rm dist}(a,b) = n} |\psi_{\alpha}(b)|^2 \to 0 \,. \nn 
\eea
A wrong scenario would be to replace $|\psi_{\alpha}(b)|^2 $ by $e^{-n/\zeta}$ with $\zeta$ the localization length previously introduced. Indeed in that case, since $\sum_{b , \,  {\rm dist}(a,b) = n} 1 \sim K^n$, one  would then conclude that (\ref{Eq:CaveatBetheLattice}) holds if and only if $\zeta < 1/\log(K)$, the localization length is bounded. This would predict a critical disorder for localization that is smaller than the one predicted before since here we have predicted a critical disorder such that the localization length diverges at the transition. The caveat here is again due to the existence of fat-tails in the distribution of $|\psi_{\alpha}(b)|^2$: in the localized phase $|\psi_{\alpha}(b)|^2$ is typically much smaller than $e^{-n/\zeta}$. In fact the sum over $b$ in (\ref{Eq:CaveatBetheLattice}) is dominated by its maximum\footnote{A more correct statement is that the sum and the maximum are of the same order. This is the `condensation' phenomenon for the sum of iid random variables $x_i$ distributed with a fat-tail $p(x)\sim_{|x| \gg 1} |x|^{-1-\alpha}$ with $\alpha \in ]0,1[$ (no first moment, as is the case here).} and  (\ref{Eq:CaveatBetheLattice}) can be replaced by 
\bea \label{Eq:CaveatBetheLattice2}
\lim_{n^* \to \infty} \sum_{n \geq n^*}  {\rm max}_{b , \, {\rm dist}(a,b) = n} |\psi_{\alpha}(b)|^2 \to 0 \,. \nn
\eea
Now $ {\rm max}_{b , \,  {\rm dist}(a,b) = n} |\psi_{\alpha}(b)|^2$ is indeed typically given by $e^{-n/\zeta}$ and \eqref{Eq:CaveatBetheLattice2} is satisfied $\forall \zeta >0$.

\section{Many-Body-Localization}  \label{sec:MBL1}

\subsection{Introduction -- Non-thermalization}  \label{subsec:MBL1:Intro}

Adding a two-body local interaction potential $U(i-j) = \frac{\lambda}{\nu} u(i-j)$ to the Anderson Hamiltonian \eqref{Eq:AndersonHamiltonian}, we now consider the Hamiltonian

\bea \label{Eq:Many-Body-Anderson}
H := - g  \left( \sum_{ \langle i j \rangle} c_i^{\dagger} c_j  + c_j^{\dagger} c_i \right) + \sum_{i } \epsilon_i c_i^{\dagger} c_i  + \frac{1}{2} \frac{\lambda}{\nu} \sum_{ij} c_i^\dagger c_j^\dagger u(i-j) c_j c_i \, ,
\eea
where $u(i-j)$ is a short-ranged dimensionless interaction kernel, $\lambda$ is a dimensionless parameter controlling the strength of the interaction, and $\nu \sim \frac{1}{W/2+2dg}$ is the density of states (per unit of volume, the relevant energy scale) of the single particle problem. Assuming that all single particle eigenfunctions $\psi_\alpha(i)$ are localized with some localization length $\xi \geq 0$ and energy $E_{\alpha}$, we rewrite (\ref{Eq:Many-Body-Anderson}) by introducing the fermionic single-particle creation (and annihilation) operators $c_\alpha^\dagger := \sum_{i} \psi_{\alpha}(i) c_i^\dagger $ as
\bea \label{Eq:Many-Body-Anderson2}
H = \sum_{\alpha} E_{\alpha} n_\alpha + \frac{\lambda}{\nu} \sum_{\alpha<\beta , \gamma < \delta} u_{\alpha\beta,\gamma \delta} c_{\alpha}^\dagger c_{\beta}^\dagger  c_{\gamma} c_{\delta} \, ,
\eea
where $n_{\alpha} = c_{\alpha}^\dagger c_{\alpha}$ is the number operator. We have assumed an arbitrary ordering between the single particle states and the interaction potential $u(i-j)$ now appears in terms of the coefficients $u_{\alpha\beta,\gamma \delta}$ which are antisymmetrized: $u_{\beta\alpha,\delta \gamma}=u_{\alpha\beta,\gamma \delta}=-u_{\beta \alpha,\gamma \delta}$ with also $u_{\alpha\beta,\gamma\delta} = u_{\gamma \delta , \alpha \beta}$. It has been argued in \cite{BAA2006,GornyiMirlinPolyakov2005} that the Hamiltonian \eqref{Eq:Many-Body-Anderson2} exhibits a phase transition as a function of $\lambda$ from a many-body-localized phase for $\lambda <\lambda_c$ to a thermal phase for $\lambda>\lambda_c$.

\paragraph{Non-thermalization} In the vaguest definition of the MBL phase, the Hamiltonian \eqref{Eq:Many-Body-Anderson2} is said to exhibit many-body-localization if the system does not thermalize under its own unitary dynamics when starting from a general many-body state $\ket{\psi}$. To make this precise let us first define thermalization: starting from a many-body state $\ket{\psi}$ (e.g. an arbitrary product state $\ket{\psi} = \prod_{i=1}^N  c^{\dagger}_{\alpha_1} \cdots c_{\alpha_N}^\dagger \ket{0}$ with $N= \nu L^d$), we say that the system thermalizes if the expectation values of arbitrary local observables $O_A$ acting only on a finite portion $A \subset \JZ^d$ of the system converges at large time to the value predicted by the Gibbs distribution. I.e. the system thermalizes iff (assuming that energy and particle number are the only conserved quantities)\footnote{That excludes $1$d integrable quantum systems where \eqref{Eq:Thermalization} is replaced by a generalized Gibbs ensemble.}
\bea \label{Eq:Thermalization}
\lim_{t \to \infty} \lim_{L \to \infty} \braket{\psi(t)| O_A |\psi(t)} = \lim_{L \to \infty} {\rm Tr} \left( O_A  \frac{e^{- \beta (H - \mu \hat N)}}{Z} \right) \, ,
\eea
with the inverse temperature $\beta$ and the chemical potential $\mu$ fixed to ensure that this holds for $O_A=H_A$ and $O_A=\hat{N}_A=\sum_{x \in A} c^\dagger_x c_x$. In the non-interacting case $\lambda = 0$, thermalization trivially breaks down but that is of course not surprising. On the other hand, it is expected on general grounds that interacting many-body systems thermalize.

One way to characterize non-thermalization is dynamical and makes the link with the characterization of Anderson localization as the absence of continuous spectrum. The statement is that if the imaginary part of the many-body self-energy associated with the retarded Green function $G_{\alpha}^R(\tau) = - i \theta(\tau) \langle \{c_{\alpha}(0) , c_{\alpha}^\dagger(\tau) \} \rangle$ vanishes when starting from a typical state and in the limit of a system that is coupled with a bath of vanishing strength, then the system does not thermalize. That can be argued by considering the quantum Boltzmann equation, and this is the point of view adopted in \cite{BAA2006,GornyiMirlinPolyakov2005}. Decay of the wave-function from sites to sites in the delocalized phase of the single particle problem is replaced in the many-body problem by the decay of non-interacting states into one another.

Another characterization can be given by considering directly the eigenstates of the many-body Hamiltonian: thermalization breaks down if these eigenstates are weak-deformations (that will be made precise later) of the non-interacting eigenstates (product of localized single particle states). The link of this statement with thermalization is made through the {\it eigenstate thermalization hypothesis} (ETH) (see \cite{Deutsch1991,Srednicki1994} and \cite{DalessioKafriPolkovnikovRigol2016} for review). This hypothesis consists in assuming that Hamiltonians of generic many-body systems are thermal if typical many-body eigenstates $\ket{\psi_\mu}$ are thermal in the sense that they satisfy Eq.~\eqref{Eq:Thermalization} with $\ket{\psi(t)} \to \ket{\psi_\mu}$. In particular, high energy eigenstates of Hamiltonian satisfying the ETH must have homogeneous profiles of particle and energy densities, and cannot therefore be close to product states. It should be stressed that the ETH is only a sufficient condition for thermalization -- the equivalence with thermalization starting from typical states has not been established. Let us also stress here that Anderson localization is in a sense a singular point of (and is included in) the MBL phase: a localized interacting system is by definition MBL, and there is to our knowledge no well-defined sense in which small interactions would be irrelevant.

\paragraph{Other properties} Other properties of the MBL phase are the absence of level repulsion and of diffusive transport, similarly to the non-interacting case. In the many-body setting, the idea that eigenstates of the system are close to product states is measured not through the calculation of some inverse participation ratio (see below) but rather through the scaling of the entanglement entropy. For an arbitrary many-body eigenstate $\ket{\psi}$ and a subsystem $A \subset \JZ^d $, one considers the reduced density matrix $\rho_A = {\rm Tr}_{{\cal H} \backslash {\cal H}_A} \left( \ket{\psi} \bra{\psi} \right)$ and the entanglement entropy $S_A = -{\rm Tr}_{{\cal H}_A} \rho_A \log (\rho_A)$. While thermal (ETH) states are expected to satisfy a volume-law $S_A \sim {\rm Vol}(A)$ (at least for eigenstates with a finite energy density above the ground state), MBL states satisfy an area law $S_A \sim {\rm Vol}(\partial A)$, similarly to ground states of many-body Hamiltonians with local interactions. In a quench protocol starting from an arbitrary product state, the measurement of the entanglement entropy of the half-chain (in the $1$d case) $A= \JZ_+$ as a function of time also displays distinctive features of the MBL phase: it grows logarithmically, $S_A(t) \sim \log(t)$, up to arbitrarily large times (in an infinite system). This last feature also distinguishes MBL from Anderson localization: in the absence of interaction the half-system entanglement entropy is bounded.

\subsection{LIOMs} \label{subsec:MBL1:LIOMs}

A characterization of the MBL phase that implies the above properties is the existence of a complete set of local integral of motions (LIOMs) $\hat{n}_\alpha$ such that the interacting Hamiltonian \eqref{Eq:Many-Body-Anderson2} can be rewritten in a classical form as
\bea \label{Eq:LIOMsHamiltonian}
H = \sum_{\alpha} \tilde{E}_{\alpha} \hat{n}_\alpha + \sum_{\alpha<\beta} J_{\alpha,\beta} \hat{n}_\alpha \hat{n}_\beta + \sum_{\alpha<\beta<\gamma} J_{\alpha,\beta,\gamma} \hat{n}_\alpha \hat{n}_\beta \hat{n}_\gamma  + \cdots \, .
\eea
The LIOMs and the expansion satisfy the following properties:
\begin{enumerate}
\item
The LIOMs are functionally independent and commute with one another and with the Hamiltonian and the number operator $\hat{N}=\sum_{x \in \JZ^d}c^\dagger_x c_x$.
\item
The LIOMs are quasi-local operators: to each $\hat{n}_\alpha$ one can associate a localization center $R_\alpha$ and given an expansion of the LIOMs $\hat{n}_\alpha = \sum_{\cI,\cJ} A_{\cI,\cJ}^{(\alpha)}O_{\cI , \cJ} $ on the basis of operators $O_{\cI , \cJ} = \prod_{x \in \cI} c_x^{\dagger} \prod_{y\in \cJ} c_y $ for $\cI$ and $\cJ$ stretches of $N$ sites $\cI = (x_1,\cdots, x_N)$ and $\cJ = (y_1,\cdots , y_N)$ one has
\bea
|A_{\cI,\cJ}^{(\alpha)}|  \leq C_\alpha \exp\left( - \frac{d(R_\alpha , \cI \cup \cJ)}{\xi_{\rm op}}  \right) \, , \nn
\eea
with $d(R_\alpha , \cI \cup \cJ)$ the distance between the localization center and the set of sites on which $O_{\cI,\cJ}$ is supported, $C_\alpha$ a constant and $\xi_{\rm op}$ the `operator localization length'.
\item
The interaction terms $J_{\alpha,\beta}$ (and higher order terms) between LIOMs also decay exponentially with the distance between the localization center, with a possibly different localization length $\xi_{{\rm int}}$.
\item
The spectrum of the LIOMs is $\{0,1\}$ and the many-body eigenstates can be labeled by their eigenvalues (this is the sense in which the set of LIOMs is complete): $\ket{\psi} = \ket{ n_j }$ with $n_j \in \{0,1\}$ and $\forall i$, $\hat{n}_i \ket{n_j } = n_i \ket{ n_j}$. There are thus $L^d$ LIOMs and $\hat{n}_i^2 = \hat{n}_i$.
\end{enumerate}
The existence of such a complete set of LIOMs has led some authors to qualify MBL as a robust form of integrability. The existence of a complete set of conservation laws is indeed a feature of integrable systems. Here the integrability is robust since the MBL phase is supposed to be a phase, while an arbitrary small perturbation of a clean integrable system breaks integrability (although that can be visible only on very long time scales). Hamiltonians that can be expressed in terms of LIOMs as \eqref{Eq:LIOMsHamiltonian} are sometimes dubbed `fully-MBL': while these do exhibit all features expected in the MBL phenomenology, the equivalence between this property and the vaguest definition of the MBL phase (non-thermalization) is not entirely clear.

\subsection{Implication of the existence of LIOMs} \label{subsec:MBL1:LIOMs2}

Here we briefly comment on the implication of the existence of a set of LIOMs; a more complete review can be found in \cite{ImbrieRosScardicchio2016review}.

 One can think of the LIOMs $\hat{n}_\alpha$ as a dressed version of the non-interacting number operators $n_\alpha$: $\hat{n}_\alpha = n_\alpha + \cdots$, with corrections vanishing as $\lambda \to 0$. This is emphasized in the non-perturbative and perturbative explicit construction of the LIOMs of \cite{Imbrie2016} and \cite{RosMullerScardicchio2015} and corresponds the point of view where MBL is thought of as localization in operator space (for numerical schemes for construction of the LIOMs see  \cite{chandran2015constructing,rademaker2016explicit,pekker2017fixed}). That can be thought of as the lattice with sites labeled by the non-interacting many-body eigenstates (product states), and connections between sites set by the interaction term in \eqref{Eq:Many-Body-Anderson2}. Note that exponential localization of eigenstates in this space does not imply $O(1)$ inverse participation ratio since the connectivity of the lattice (and thus the number of points-states at a given distance of a point) also scale with $N=L^d$. On the other hand the quasi-locality of the LIOMs implies that eigenstates follow an area law for the entanglement entropy. The proximity of true eigenstates to product state also implies a breaking of the ETH. The exponential decay of the interaction terms between LIOMs has been argued to lead to the characteristic logarithmic growth of the half-chain entanglement entropy after a quench to a typical high energy state. Finally, the existence of LIOMs implies a vanishing conductivity and thus the absence of diffusive transport. This can be shown\cite{RosMullerScardicchio2015} by considering the Kubo formula for the real part of the DC conductivity at inverse temperature\footnote{Here the use of the canonical ensemble at inverse temperature $\beta$ in \eqref{Eq:Kubo} is for the sake of concreteness. It could be replaced by any other probability distribution on the eigenstates of the Hamiltonian, the Kubo formula still remaining valid anyway.} $\beta>0$
\be \label{Eq:Kubo}
\Re(\sigma(w=0))= \lim_{\eta \to 0} \lim_{L \to \infty} \frac{\pi \beta}{L^d} \sum_{r'r} \sum_{m\neq m'} \frac{e^{-\beta E_{m'}}}{Z} \braket{E_{m'}|J_{r'+r}|E_m} \braket{E_{m}|J_{r'}|E_{m'}} \delta_{\eta}(E_{m'}-E_m) \, ,
\ee
where $Z$ is the partition sum, $\ket{E_m}$ are the many-body eigenstates of the system, $J_r$ is the current operator that acts only on a few spins around $r$, and $\delta_{\eta}(E) =  \frac{1}{\pi} \frac{\eta}{\eta^2+E^2}$ is a smoothed delta function. Since the LIOMs are complete, for any pair of many-body eigenstates $(\ket{E_m},\ket{E_{m'}})$ there is at least one LIOM $\hat{n}_\alpha$ which differ on these states: $\hat{n}_\alpha \ket{E_m} = n_\alpha^m \ket{E_m}$ and $\hat{n}_\alpha \ket{E_{m'}} = n_\alpha^{m'} \ket{E_{m'}}$ with $n_\alpha^m \neq n_\alpha^{m'}$. Using that LIOM, the matrix element in \eqref{Eq:Kubo} are rewritten as
\bea
\braket{E_{m'}|J_{r'+r}|E_m} = \frac{\braket{E_{m'}|[J_{r'+r} , \hat{n}_\alpha] | E_m}}{n_\alpha^m-n_\alpha^{m'}} \quad , \quad \braket{E_{m}|J_{r'}|E_{m'}} = \frac{\braket{E_{m}|[J_{r'} , \hat{n}_\alpha] | E_{m'}}}{n_\alpha^{m'}-n_\alpha^{m}} \, . \nn
\eea
If the LIOMs are strictly local with a support of size $\xi_{{\rm op}}$ (i.e. in the absence of exponential tail), one of the two matrix elements is $0$ for $r \geq \xi_{{\rm op}}$ and the sum over $r$ in \eqref{Eq:Kubo} can thus be restricted to $r \geq \xi_{{\rm op}}$. Since $J_{r'}$ is local $J_{r'} \ket{m}$ can only differ from $\ket{m}$ by a finite number of LIOMs and the sums over $m'$ in \eqref{Eq:Kubo} is also restricted to a finite number of terms. This shows that the limit $\lim_{\eta \to 0}$ in \eqref{Eq:Kubo} can be taken inside the sum and the $\delta_{\eta}(E_{m'}-E_m)$ then vanishes with probability $1$. In the case of quasi-local LIOMs, the presence of exponential tails makes this argument insufficient as the exponential decay of $\braket{E_{m}|J_{r'}|E_{m'}}$ for states $\ket{E_{m'}}$ differing from $\ket{E_m}$ by LIOMs that are located at a distance $\ell$ from $r'$ can in principle compete with the exponential decay of the many-body level spacing $\sim e^{- 2 \ell^d  \log(2)}$ (at least in $d=1$) and lead to a finite conductivity. This connects with recent bounds obtained on the `typical' LIOMs localization length, see \cite{RoeckHuveneers2017,LuitzHuveneersRoeck2017,ThieryHuveneersMullerRoeck2017}. Let us finally mention here that in $d=1$ the absence of diffusive transport is also observed in the thermal phase when sufficiently close to the critical point, when rare localized regions (Griffiths regions) of the material can serve as bottlenecks and lead to subdiffusive transport, see \cite{AgarwalGopalakrishnanKnapMullerDemle2015r,GopalakrishnanAgarwalDemlerHuseKnap2016,VoskHuseAltman2015,BarLevCohenRechman2015,vznidarivc2016diffusive} and \cite{AgarwalAltmanDemlerGopalakrishnanHuseKnap2017} for a review.

\subsection{Spin systems}  \label{subsec:MBL1:Spins}

The idea of a non-thermalization transition in the form of an MBL/ETH transition applies well beyond the scope of interacting electrons in disordered solids and other strongly disordered many-body quantum systems, in particular spin systems, can be considered\footnote{The role of disorder has also been questioned, and although it is now increasingly clear that the fully-MBL phenomenology requires disorder \cite{DeRoeckHuveneers2014b}, it is also clear that translation invariant disorder can exhibit similar features (extremely slow transport), sometimes dubbed `asymptotic MBL' \cite{DeRoeckHuveneers2014b,DeRoeckHuveneers2014a,SchiulazMuller2014,GroverFisher2014}.}. A prototypical example is a random transverse field Ising model
\bea
H_{{\rm spin}} = \sum_i h_i S_i^z + \sum_i \gamma_i S_i^x + \sum_{i}{} J_i S_i^z S_{i+1}^z \, , \nn 
\eea
with random field $h_i \sim h$, transverse field $\gamma_i \sim \gamma$, interaction term $J_i \sim J$ and $S_i^\alpha$ the Pauli matrices. Here for $\gamma = 0$ the Hamiltonian is diagonal in the $S_i^z$ basis and for $\gamma << h,J$ it has been suggested that the system exhibits MBL in the same sense as before: the Hamiltonian can be rewritten in the classical form
\bea
H_{{\rm spin}} = \sum_i \tilde{h}_i \tau_i^z + \sum_i \tilde{J}_{i,i+1} \tau_i^z \tau_{i+1}^z + \sum_i \tilde{J}_{i,i+1,i+2} \tau_i^z \tau_{i+1}^z \tau_{i+2}^z + \cdots \, , \nn
\eea
with the LIOMs $\tau_i^z$ quasi-local operators that are dressed version of the $S_i^z$. In that context the LIOMs are often called l-bits (logical) while the original spin $S_i^z$ are called p-bits (physical). This has been (up to an assumption) proved explicitly in $d=1$ in \cite{Imbrie2016} where the LIOMs are explicitly constructed as $\tau_i^z=U S_i^z U^\dagger$ with $U$ an unitary operators that preserves the locality and that is close to the identity almost everywhere for $\gamma \ll h,J$. 

\section{Forward Scattering Approximation approach to MBL} \label{sec:MBL2}

In this section we review the forward scattering approximation approach to MBL proposed in \cite{RosMullerScardicchio2015}. Pedagogical reviews of this approach can also be found in \cite{ImbrieRosScardicchio2016review,RosPHD}.

\subsection{A coarse grained model} \label{subsec:MBL2:model}

We consider the interacting model (\ref{Eq:Many-Body-Anderson2}) that we recall here for readability:
\bea \label{Eq:MBHamRecall}
&& H = H_0+ \lambda U  \quad , \quad H_0 = \sum_{\alpha} E_{\alpha} n_\alpha \quad , \quad U=  \frac{1}{\nu} \sum_{\alpha<\beta , \gamma < \delta} u_{\alpha\beta,\gamma \delta} c_{\alpha}^\dagger c_{\beta}^\dagger  c_{\gamma} c_{\delta} \, .
\eea
The model is corse-grained as follows. We assume that the singe particle localization length $\xi$ is large, $\xi \gg 1$, and simplify the interaction terms by grouping single particle states into `localization volumina' of size $\xi^d$ and assuming that $u_{\alpha \beta, \gamma \delta}$ is non-zero iff: (i) all four states are inside the same are adjacent localization volumina; (ii) the energy of the the states are `close enough':
\bea
|E_\alpha - E_\delta|, |E_{\beta} - E_{\gamma}| \leq \delta_\xi \quad \text{or} \quad |E_\alpha - E_\gamma|, |E_{\beta} - E_{\delta}| \leq \delta_{\xi} \, ,  \nn 
\eea
where the energy scale $\delta_\xi$ is the typical level spacing of single particle states inside the same localization volume:
\bea
\delta_\xi \sim \frac{1}{\nu \xi^d} \, . \nn 
\eea
In that case we assume that 
\bea
u_{\alpha \beta , \gamma \delta}= \nu \delta_\xi \eta_{\alpha \beta , \gamma \delta} \, , \nn 
\eea
with the $\eta_{\alpha \beta , \gamma \delta}$ some $O(1)$ iid random variables, taken for simplicity as uniformly distributed in $[-1,1]$. The largeness of $\xi$ ensures that there is a large number $N_{{\rm loc}} \sim \xi^d$ of wave-function in each localization volumina. This will later on provide a large parameter that will simplify the analysis. Setting $u_{\alpha \beta , \gamma \delta} \sim \nu \delta_\xi$ ensures that the interaction efficiently hybridize the states that are the closest in energy when the dimensionless parameter $\lambda$ is of order $1$, while ignoring the interaction between levels that are separated by larger energy is natural since these will then be much less likely to be hybridized by the interaction.

\subsection{Naive and renormalized perturbation theory} \label{subsec:MBL2:PerturbationTheory}

\paragraph{Naive perturbation theory}
In the Anderson problem, the most natural idea to construct the localized eigenstates was to perform a naive perturbation theory in the coupling. Similarly here one tries to construct the LIOMs $\hat{n}_{\alpha}$ using a perturbation theory in the interaction term, i.e. in $\lambda$. We thus write
\bea
\hat{n}_{\alpha} = n_\alpha + \sum_{n\geq 1} \lambda^n \Delta \hat{n}_{\alpha}^{(n)} . 
\eea
And $[H,\hat{n}_{\alpha}] = 0$ implies the recursion equation, for $n \in \JN$ (writing $ \Delta \hat{n}_{\alpha}^{(0)} =n_\alpha $)
\bea \label{Eq:LIOMSRec1}
[H_0 ,\Delta \hat{n}_{\alpha}^{(n+1)} ] = - [U, \Delta \hat{n}_{\alpha}^{(n)}] \, .
\eea 
This is a recursion equation in the space of particle-conserving operators ($[\hat{n}_\alpha , \hat N] = 0$ is enforced at every order) that we denote ${\cal C}$. Eq.~\eqref{Eq:LIOMSRec1} is not invertible since the map $Ad_{{H_0}}(X) = [H_0, X]$ is obviously not a bijection. Writing $K$ the kernel of the map and $O=[H_0, {\cal C}]$ which span ${\cal C}$ as ${\cal C} = K \oplus O$ ($K$ and $O$ are orthogonal spaces for the Hermitian product $\langle A,B \rangle:={\rm Tr}[A^\dagger B]$), this means that at every order $\Delta \hat{n}_{\alpha}$ is given by 
\bea
\Delta \hat{n}_{\alpha}^{(n)} = \Delta J_\alpha^{(n)} + \Delta K_\alpha^{(n)} \, , \nn 
\eea
where $\Delta J_\alpha^{(n)} \in O$ is thus uniquely determined by (\ref{Eq:LIOMSRec1}). An explicit solution for $\Delta J_\alpha^{(n)}$ can in fact be written as
\bea
\Delta J_\alpha^{(n+1)} = i \lim_{\eta \to 0} \int_0^\infty d\tau e^{-\eta \tau} e^{i \tau H_0}[U , \Delta \hat{n}_{\alpha}^{(n)}] e^{-i \tau H_0} \, .
\eea
On the other hand the remaining term $\Delta K_\alpha^{(n)}$ can be uniquely fixed by requiring that the LIOMs satisfy $\hat{n}_\alpha^2 = 1$. In the following we will only construct the operators
\bea 
I_{\alpha} := n_\alpha + \sum_{n\geq 1} \lambda^n \Delta J_{\alpha}^{(n)} \, , 
\eea
that are thus equal to the LIOMs up to a `normalization'. The goal is to check that these operators can indeed be constructed and are quasi-local.  It is natural to assume that if this is the case then the LIOMs can also be constructed and are quasi-local, implying MBL.

\paragraph{Renormalized perturbation theory} As in the one-particle case, this naive perturbation theory in $\lambda$ can only diverge in the thermodynamic limit due to the presence of local resonances. The expansion is reformulated as
\bea \label{Eq:ExpansionLIOMsI}
I_{\alpha} = n_{\alpha} + \sum_{N \geq 1} \sum_{\cI \neq \cJ, |\cI|=|\cJ|} \cA_{\cI, \cJ}^{(\alpha)} \left( \cO_{\cI , \cJ} +  \cO_{\cI , \cJ}^{\dagger} \right)
\eea
where $\cI$ and $\cJ$ are multi-indices: $\cI = (\beta_1,\cdots, \beta_N)$ and $\cJ = (\gamma_1,\cdots , \gamma_N)$ and
\bea
\cO_{\cI , \cJ} := \prod_{\beta \in \cI} c_\beta^{\dagger} \prod_{\gamma \in \cJ} c_\gamma 
\eea
 is a normally ordered operator. This expansion is coherent with the precedent formal definition of $I_{\alpha}$ since $I_{\alpha}-n_{\alpha}$ was introduced as an element of $O= [H_0,{\cal C}]$. Indeed, note that ${\cal C} = O \oplus K$ with $K$ the kernel of the map $Ad_{H_0}$, which is exactly the ensemble of operators $\cO_{\cI, \cJ}$ with $\cI=\cJ$. Being restricted to $\cI \neq \cJ$, the expansion (\ref{Eq:ExpansionLIOMsI}) indeed defines $I_{\alpha}$ such that $I_{\alpha} - n_{\alpha} \in O$. This also implies that the coefficients $A_{\cI, \cJ}^{(\alpha)}$ are uniquely defined by the equation $[H,I_{\alpha}]=0$. The fact that the equation is well-posed, and thus always admits a solution in a finite system, will be shown below. Note finally that (\ref{Eq:ExpansionLIOMsI}) is not thought of as an expansion in $\lambda$ anymore. Even though this is perhaps less apparent here than in the Anderson case where self-energies were introduced, this expansion is the equivalent of the renormalized perturbation theory of the singe particle case. In the infinite system, the existence of a solution of the form (\ref{Eq:ExpansionLIOMsI}) is equivalent to the convergence of the expansion (\ref{Eq:ExpansionLIOMsI}), which can then be reinterpreted as an expansion in the support of the operators $\cO_{\cI , \cJ}$, defined as $r(\cI,\cJ):={\rm max}_{(\beta,\gamma) \in \cI \times \cJ} |r_{\beta} - r_{\gamma}|$ with $r_{\alpha}$ the localization center of the single particle state $\ket{\alpha}$. The convergence of the expansion is then interpreted as in the Anderson case as a signature of MBL. We thus want to show that, $\forall \epsilon> 0$
\bea \label{Eq:ExpansionLIOMsConvergence}
\lim_{R \to \infty} \mathbb{P}\left( \sum_{\cI \neq \cJ, \,  r(\cI ,\cJ)>R}  |\cA_{\cI, \cJ}^{(\alpha)}| < \epsilon \right) =1 \, .
\eea
That will be shown below by computing the $\cA_{\cI, \cJ}$ at the lowest order in $\lambda$ (forward approximation), which, as in the Anderson case, defines an expansion whose convergence properties are {\it a priori} better than the naive perturbation theory, but less good than the renormalized expansion (see however \cite{DeRoeckHuveneersMullerSchiulaz2016,RoeckHuveneers2017} for a discussion on the role of rare events not taken into account here).

\subsection{Localization of LIOMs as single particle localization in operator space} \label{subsec:MBL2:ALinOpSpace}

\paragraph{Recursion equations for the amplitudes} Imposing $[H,I_{\alpha}] = 0$ leads to the following recursion equation for the amplitudes $\cA_{\cI, \cJ}^{(\alpha)}$, noting  $\cI=(\alpha_1,\cdots,\alpha_N)$ and $\cJ=(\beta_1,\cdots , \beta_N)$ with $2 N \geq 4$ the number of creation and annihilation operators in $\cO_{\cI,\cJ}$ ($N$ is also called `level' in the following) 
\bea  \label{Eq:LIOMSRecA}
&& 0=  \left(  \sum_{n=1}^N \frac{E_{\alpha_n}-E_{\beta_n}}{\delta_{\xi}}  \right)\cA_{\cI, \cJ}^{(\alpha)}  \nn \\
&&+ \lambda \sum_{1 \leq l<m \leq N} \left[ \sum_{\gamma < \delta}\left( \eta_{\alpha_l \alpha_m , \gamma \delta} \tilde{\cA}_{\cI_{lm}^{\gamma \delta}, \cJ}^{(\alpha)} -\eta_{\gamma \delta,\beta_l \beta_m} \tilde{\cA}_{\cI, \cJ_{lm}^{\gamma \delta}}^{(\alpha)}   \right)    \right] \nn \\
&& + \lambda \sum_{1 \leq l<m \leq N} \sum_{n=1}^N (-1)^{N+1} \left[ \sum_{\gamma}\left( \eta_{\alpha_l \alpha_m , \gamma \beta_n} \tilde{\cA}_{\cI_{lm}^{\gamma }, \cJ_n}^{(\alpha)} -\eta_{\gamma \alpha_n,\beta_l \beta_m} \tilde{\cA}_{\cI, \cJ_{lm}^{\gamma}}^{(\alpha)}   \right)    \right] \, ,
\eea
where we have introduced several shorthands for any index set $\cchi=(x_1,\cdots,x_N)$: (i) the diagonal elements $\cA_{\cchi,\cchi}^{(\alpha)}$ are $0$ except if $\cchi=\alpha$: $\cA_{\cchi,\cchi}^{(\alpha)} = \delta_{\cchi,\{\alpha\}}$, allowing to include easily the `initial condition' in (\ref{Eq:LIOMSRecA}); (ii) $\forall l,m$ indices with $1\leq l < m \leq N$ and single particle labels $\gamma,\delta$, $\cchi_{lm}^{\gamma,\delta} := (\gamma , \delta , x_1 , \cdots , \cancel{x_l} , \cdots , \cancel{x_m},\cdots , x_N)$ and similarly for $\cchi_{lm}^{\gamma} $ etc; (iii) The resulting created index sets $\cchi'$ being not ordered (with respect to the single particle label), we define $s[\cchi']$ the signature of the permutation that orders $\cchi'$ and in general $ \tilde{\cA}_{\cI , \cJ}:=(-1)^{s[\cI]+s[\cJ]} \cA_{\cI , \cJ}$, thus taking care of the fermionic commutations relations.

\medskip

In the equation $\frac{1}{\delta_\xi} [H,I_{\alpha}]= \frac{1}{\delta_\xi}[H_0 ,I_{\alpha}] +  \lambda [\frac{U}{\delta_{\xi}},I_{\alpha}] = 0$, the first line of (\ref{Eq:LIOMSRecA}) corresponds to $[H_0 ,I_{\alpha}]$, while the second and the third correspond to $  \lambda [\frac{U}{\delta_{\xi}},I_{\alpha}]$ and to two types of scattering processes. Indeed, for the first line write 
\bea
[c_{\alpha}^{\dagger}c_{\alpha},c_{\alpha_1}^{\dagger}\cdots c_{\alpha_N}^{\dagger} c_{\beta_1}\cdots c_{\beta_N}]  && = c_{\alpha}^{\dagger}c_{\alpha} c_{\alpha_1}^{\dagger}\cdots c_{\alpha_N}^{\dagger} c_{\beta_1}\cdots c_{\beta_N}-c_{\alpha_1}^{\dagger}\cdots c_{\alpha_N}^{\dagger} c_{\beta_1}\cdots c_{\beta_N} c_{\alpha}^{\dagger}c_{\alpha}  \nn \\
&& = (\sum_{i=1}^N  \delta_{\alpha_i , \alpha}  - \sum_{i=1}^N  \delta_{\beta_i , \alpha}) c_{\alpha_1}^{\dagger}\cdots c_{\alpha_N}^{\dagger} c_{\beta_1}\cdots c_{\beta_N} \, . \nn 
\eea
For the second and third, write
\bea
&& [c_{\alpha}^{\dagger}c_{\beta}^{\dagger}c_{\gamma}c_{\delta},c_{\alpha_1}^{\dagger}\cdots c_{\alpha_N}^{\dagger} c_{\beta_1}\cdots c_{\beta_N}]   = c_{\alpha}^{\dagger}c_{\beta}^{\dagger}c_{\gamma}c_{\delta} c_{\alpha_1}^{\dagger}\cdots c_{\alpha_N}^{\dagger} c_{\beta_1}\cdots c_{\beta_N}-c_{\alpha_1}^{\dagger}\cdots c_{\alpha_N}^{\dagger} c_{\beta_1}\cdots c_{\beta_N}c_{\alpha}^{\dagger}c_{\beta}^{\dagger}c_{\gamma}c_{\delta}  \nn \\
&& = \left\{ \sum_{i=1}^N (-1)^{N+i}  \left( \delta_{\alpha_i,\gamma} c_{\alpha_1}^{\dagger}\cdots \cancel{c_{\alpha_i}} \cdots c_{\alpha_N}^{\dagger}c_{\alpha}^{\dagger}c_{\beta}^{\dagger}c_{\delta}+ \delta_{\alpha_i,\delta} c_{\alpha_1}^{\dagger}\cdots \cancel{c_{\alpha_i}} \cdots c_{\alpha_N}^{\dagger}c_{\alpha}^{\dagger}c_{\beta}^{\dagger}c_{\gamma}  \right) + \sum_{i\neq j =1}^N (-1)^{i+j} \right.   \nn \\
&& \!\!\!\!\!\!\!\!\!\!\!\!\!\!\!\!\!\!\!\! \left.   \delta_{\alpha_i,\delta} \delta_{\alpha_j,\gamma} c_{\alpha_1}^{\dagger}\cdots \cancel{c_{\alpha_i}}\cdots \cancel{c_{\alpha_j}} \cdots c_{\alpha_N}^{\dagger}c_{\alpha}^{\dagger}c_{\beta}^{\dagger}  \right\} c_{\beta_1}\cdots c_{\beta_N} -c_{\alpha_1}^{\dagger}\cdots c_{\alpha_N}^{\dagger} \left\{ \sum_{i=1}^{N}(-1)^i \left( \delta_{\beta_i,\alpha} c_{\beta}^{\dagger}c_{\gamma}c_{\delta} c_{\beta_1}\cdots \cancel{c_{\beta_i}} \cdots c_{\beta_N}   \right. \right. \nn \\
&&  \left. \left.  + \delta_{\beta_i,\beta} c_{\alpha}^{\dagger}c_{\gamma}c_{\delta} c_{\beta_1}\cdots \cancel{c_{\beta_i}} \cdots c_{\beta_N}   \right)  + \sum_{i\neq j =1}^N (-1)^{i+j} \delta_{\beta_i,\alpha} \delta_{\beta_j,\beta}  c_{\gamma}c_{\delta} c_{\beta_1}\cdots \cancel{c_{\beta_i}} \cdots \cancel{c_{\beta_j}} \cdots c_{\beta_N}     \right\}  \, . \nn
\eea 
And the third line of (\ref{Eq:LIOMSRecA}) corresponds to the above terms that increase the number of creation and annihilation operators by $1$, corresponding to a process where exactly one fermion/hole scatter on a hole/fermion and an additional particle-hole pair gets created, while the second line of (\ref{Eq:LIOMSRecA}) corresponds to the above terms that conserve the number of creation and annihilation operator, corresponding to a process where two fermions/holes scatter on two holes/fermions.

\begin{figure}
\centerline{\includegraphics[width=16cm]{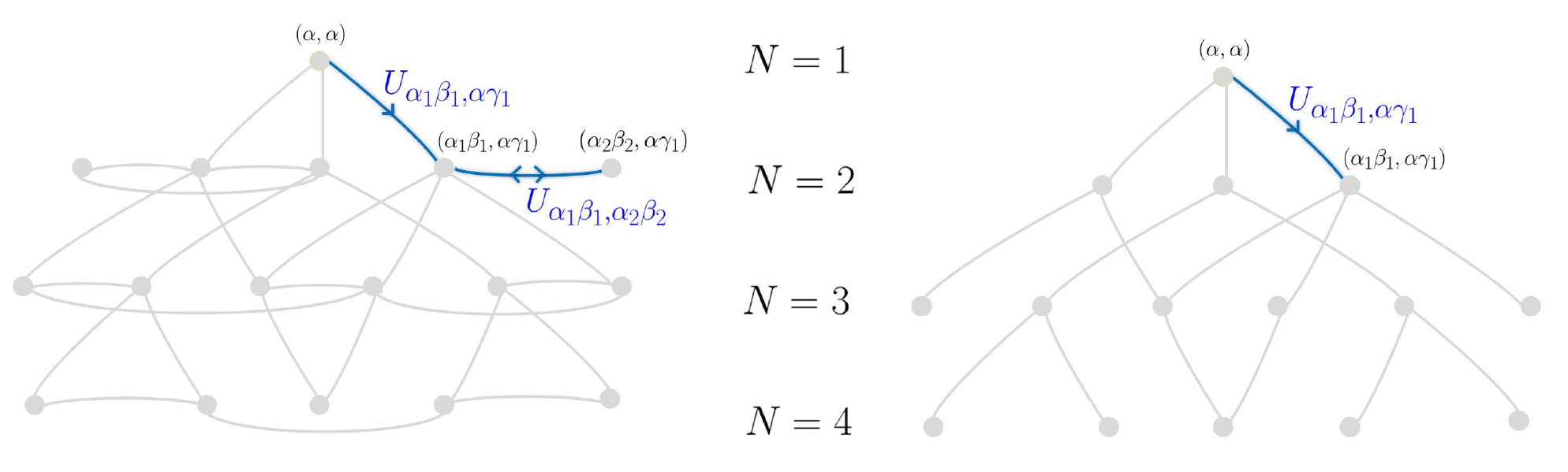}} 
\caption{Left: the perturbative construction of the LIOMs is equivalent to a non-Hermitian Anderson localization problem in the operator space, with sites labeled by multi indices $(\cI,\cJ)$, and links set up by the interaction as in \eqref{Eq:LIOMSRecA}. Right: in the forward approximation, only interaction terms linking Fock indices at level $N$ to Fock indices at level $N+1$ are retained. Figure taken from \cite{RosMullerScardicchio2015}.}
\label{Fig:operatorspace}
\end{figure} 

\paragraph{A hopping problem in operator space} Eq.~(\ref{Eq:LIOMSRecA}) shows that one can think of the amplitudes $\cA^{(\alpha)}_{\cI,\cJ}$ as the amplitude of an eigenfunction (with eigenvalue $0$) for an Anderson model on the `operator lattice' whose sites are labeled by the $(\cI , \cJ)$, whose connectivity is set by the interaction and where the on-site disorder is $\cE_{\cI,\cJ} = \sum_{\alpha \in \cI} E_{\alpha}-\sum_{\beta \in \cJ} E_{\beta} $ (and is thus correlated). This is a non-Hermitian hopping problem since the hopping never goes from level $N+1$ to $N$ (see Fig.~\ref{Fig:operatorspace}).

\subsection{Forward approximation} \label{subsec:MBL2:FWA}

\paragraph{Main approximation} To go further we now uses the forward approximation, which in this context consists in neglecting in (\ref{Eq:LIOMSRecA}) the interaction terms connecting amplitudes at the same level $N$ (second line of (\ref{Eq:LIOMSRecA})). This is partially justified (see \cite{RosMullerScardicchio2015} for a more complete discussion) by the fact that most links leaving a given site $(\cI , \cJ)$ on the $N$th level are links leading to the $(N+1)$th level. Indeed, links from level $N$ to $N+1$ are defined by (i) choosing a particle or a hole $\alpha$ in $(\cI,\cJ)$ ($2N$ choices) which can scatter on the hole or the particle that is the closest in energy in the same localization volumina ($2$ choices); (ii) choosing a new particle hole-pair in the same or in one of the adjacent localization volumina ($3\times2\times N_{{\rm loc}}$) choices. The number of these links is thus typically $3{\cal K}$ with 
\bea 
{\cal K}:= 4 N_{{\rm loc}} \gg 1 \, , \nn 
\eea
while there are only $O({\cal K}^0)$ links from level $N$ to $N$. 

That implies that at level $N$ the amplitudes $\cA_{\cI, \cJ}^{(\alpha)}$ which are linked to $(\alpha,\alpha)$ on the operator lattice by a path of length $N-1$, and are thus exactly $O(\lambda^{N-1})$ (see Fig.~\ref{Fig:operatorspace}). The solution of (\ref{Eq:LIOMSRecA}) can then be written as a sum over directed polymer paths on the operator lattice: for $\cA^{(\alpha)}_{\cI,\cJ}$ at the $N$th level we have
\bea \label{Eq:cADP}
\cA^{(\alpha)}_{\cI,\cJ} = \sum_{{\rm DP} \,  \pi: (\alpha,\alpha) \to (\cI, \cJ)} w(\pi) \quad , \quad w(\pi) := (-1)^{\sigma_{{\rm path}}} (\lambda)^{N-1} \prod_{t=0}^{N-2} \frac{\eta_{\pi(t) \to \pi(t+1)}}{\cE_{\pi(t)}} \, ,
\eea
where the $\eta_{\pi(t) \to \pi(t+1)}$ is the interaction term that was used in the transition from $\pi(t) = (\cI(t),\cJ(t)) \to \pi(t+1) = (\cI(t+1),\cJ(t+1)) $ and $(-1)^{\sigma_{{\rm path}}}$ is a global sign resulting from the fermionic commutation relations. Note that the energy encountered along the way $\cE_{\pi(t)}$ are strongly correlated along a path and between the different paths (many paths indeed connect $(\alpha,\alpha)$ to a given $ (\cI,\cJ)$, see below).

\paragraph{Localization criterion in the FWA}  At this level of approximation the series (\ref{Eq:ExpansionLIOMsI}) effectively becomes a series in $\lambda^{N-1}$. For $(\cI,\cJ)$ at the level $N$, due to the locality of the interaction, the distance $r(\cI,\cJ)$ is bounded as  $r(\cI,\cJ) < N \zeta$. A sufficient criterion to ensure the convergence of the expansion (and MBL) (\ref{Eq:ExpansionLIOMsConvergence}) is that there is $z \in ]0,1[$ such that
\bea \label{Eq:ExpansionLIOMsConvergenceN}
\lim_{N^* \to \infty} \mathbb{P}\left( \forall N >N^*, \sum_{\cI\neq \cJ , \, |\cI|=|\cJ|=N +1}  |\cA_{\cI,\cJ}^{(\alpha)}|  < z^N\right) = 1 \, .
\eea
Due to the presence of a fat tail in the distribution of $|\cA_{\cI,\cJ}^{(\alpha)}|$ (as in the Anderson problem this is a consequence of the presence of energy denominators in the expression for $|\cA_{\cI,\cJ}^{(\alpha)}|$) one expects the sum over amplitudes to be of the same order as the largest amplitude. We will thus rather investigate that
\bea \label{Eq:ExpansionLIOMsConvergenceN2}
\lim_{N^* \to \infty} \mathbb{P}\left( \forall N >N^*, {\rm max}_{\cI\neq \cJ , \, |\cI|=|\cJ|=N +1}  |\cA_{\cI,\cJ}^{(\alpha)}|  < z^N\right) = 1 \, .
\eea
Still, this probability is difficult to calculate since the different amplitudes are strongly correlated. Doing as if they were independent we need to calculate
\bea \label{Eq:ExpansionLIOMsConvergenceN3}
\lim_{N^* \to \infty}  \prod_{N \geq N^*} \mathbb{P}\left({\rm max}_{\cI\neq \cJ , \, |\cI|=|\cJ|=N +1}  |\cA_{\cI,\cJ}^{(\alpha)}|  < z^N\right) = 1 \, .
\eea
Finally, below we will show that for $\lambda$ small enough in the localized phase $ \mathbb{P}\left({\rm max}_{\cI\neq \cJ , |\cI|=|\cJ|=N +1}  |\cA_{\cI,\cJ}^{(\alpha)}|  < z^N\right)$ converges to $1$ as $e^{-e^{-N C} } $ with $C>0$ (as was found in the FWA approach to Anderson localization on the Bethe lattice). This fast convergence is then sufficient to ensure (\ref{Eq:ExpansionLIOMsConvergenceN3}). We will thus focus on showing the criterion
\bea \label{Eq:ExpansionLIOMsConvergenceN4}
\lim_{N\to \infty}  \mathbb{P}\left({\rm max}_{\cI\neq \cJ , \, |\cI|=|\cJ|=N +1}  |\cA_{\cI,\cJ}^{(\alpha)}|  < z^N\right) = 1 \, ,
\eea
which is now equivalent to the localization criterion used in the FWA approximation for Anderson localization (\ref{Eq:LocCriterion1}), and we may identify the smallest value of $z$ such that the above holds as the inverse localization length of the localization problem on the operator lattice. Note that this discussion is reminiscent of the remark on the localization criterion in the end of Sec.~\ref{subsubsec:AL2:FWABethe}.

\subsection{Effective paths} \label{subsec:MBL2:EffectivePaths}

In the following we will evaluate the probability (\ref{Eq:ExpansionLIOMsConvergenceN4}) assuming that some paths appearing in the decomposition of the amplitude  $\cA_{\cI,\cJ}$ (see \eqref{Eq:cADP}) are independent. One cannot however simply assume that {\it all} paths in (\ref{Eq:cADP}) are independent. Indeed, there is a superexponential ($\sim$ factorial) number (in $N$) of such paths, and it will become clear later on that assuming that all paths are independent would lead to delocalization. This is a consequence of the fact that the probability to observe a path of length $N+1$ with a weight $w(\pi)$ that is larger that $z^N$, i.e. $\mathbb{P}(|w(\pi)| < z^N)$, only decays exponentially in $N$ and this decay can only compete with an exponentially increasing number of terms, as was also the case in the FWA treatment of Anderson Localization on the Bethe lattice (see Sec.~\ref{subsubsec:AL2:FWABethe}).

\begin{figure}
\centerline{\includegraphics[width=8cm]{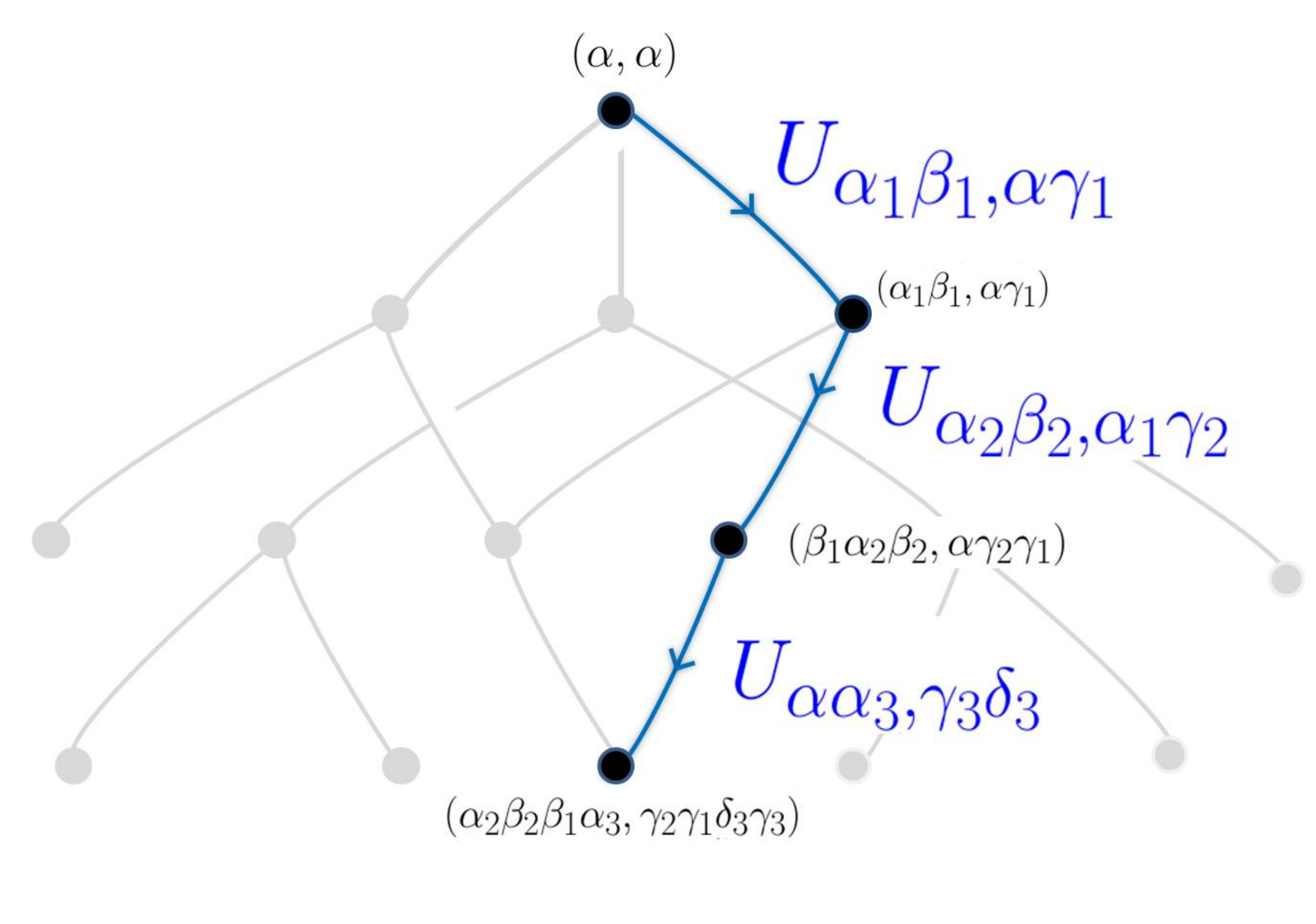} \includegraphics[width=8cm]{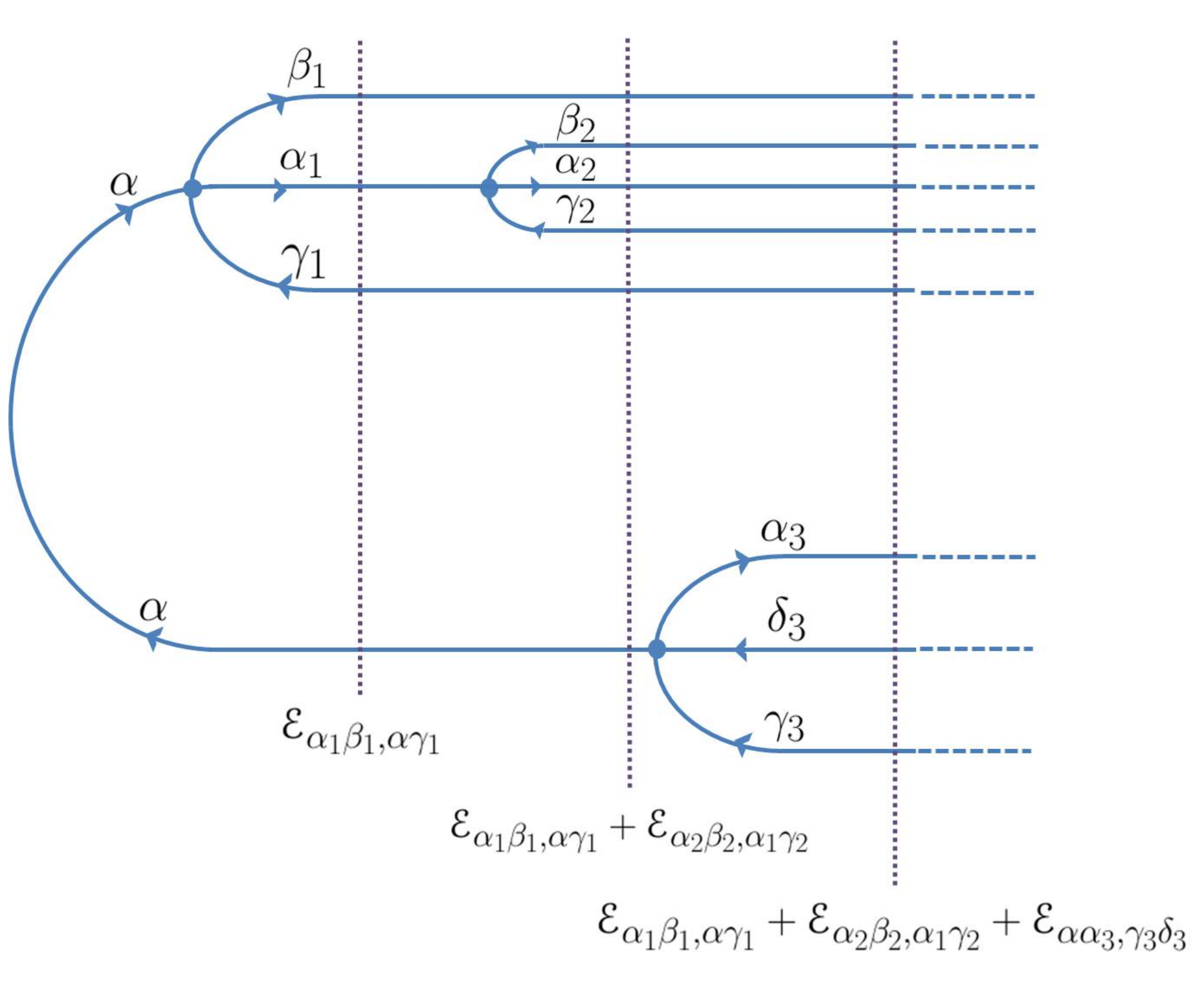}} 
\caption{A directed path on the operator space from $(\alpha,\alpha)$ to $(\cI,\cJ)$ (left), corresponding to one contribution to the amplitude $\cA^{(\alpha)}_{\cI,\cJ}$ (see \eqref{Eq:cADP}), may be represented as on the right where the scattering events are ordered from left to right. On the right a right (left) going arrow represents a creation (annihilation) operator). Figure taken from \cite{RosMullerScardicchio2015}.}
\label{Fig:diagrams1}
\end{figure} 

\begin{figure}
\centerline{\includegraphics[width=8cm]{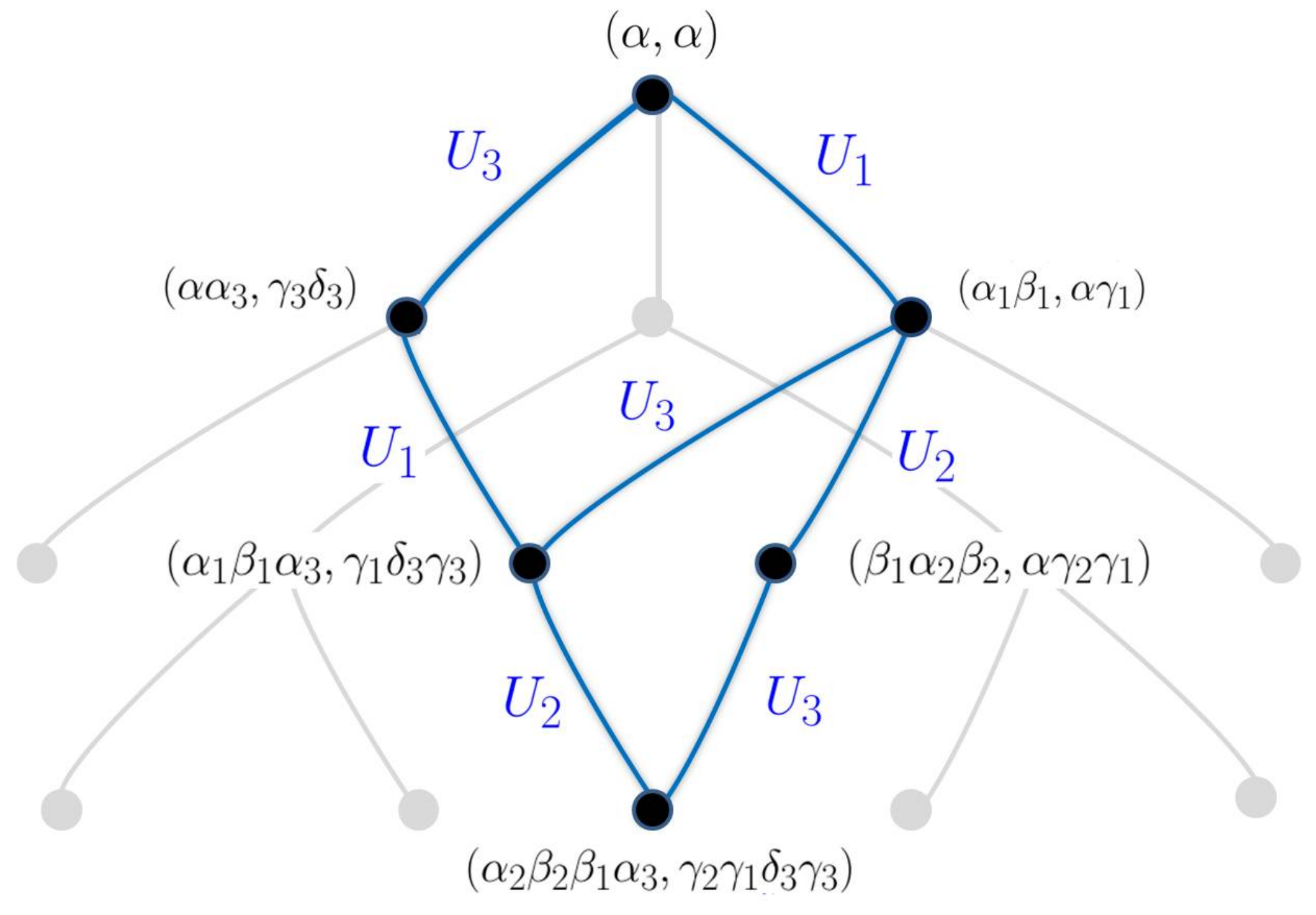}}
\vspace{1cm}
\centerline{\includegraphics[width=16cm]{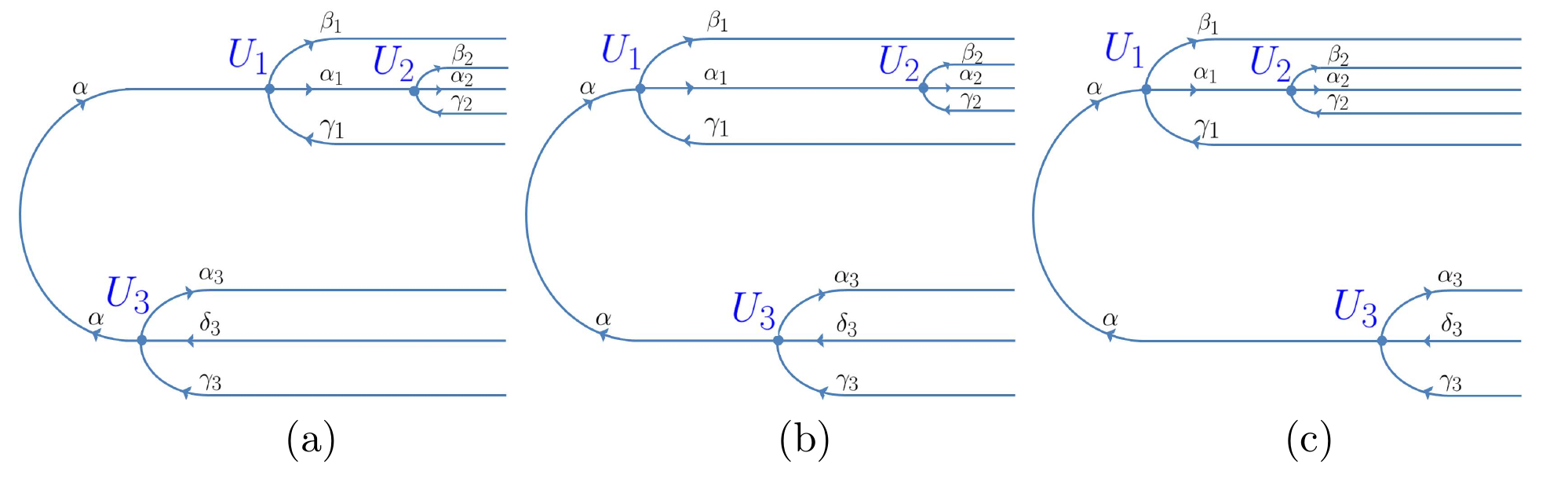}} 
\caption{Paths $\pi$ connecting $(\alpha,\alpha)$ to $(\cI,\cJ)$ that differ only by an ordering of the scattering event are regrouped into diagrams $d$. Due to the correlations between these paths, the sum of amplitudes $w(\pi)$ is reduced to a sum over a smaller number of effective amplitudes $\tilde{w}(\pi)$. Figure taken from \cite{RosMullerScardicchio2015}.}
\label{Fig:effectivepaths}
\end{figure}

\paragraph{Paths, diagrams and effective paths} A way to deal with this `factorial problem' is to resum some families of highly correlated paths. Representing graphically a path as in Fig.~\ref{Fig:diagrams1} we regroup all paths that differ from one another only by a change in the ordering of the interaction (see e.g. Fig.~\ref{Fig:effectivepaths}). Such a family of paths defines a {\it diagram $d$}, and we define $\cD_{\cI,\cJ}$ the ensemble of diagrams going from $(\alpha,\alpha)$ to $(\cI,\cJ)$. The amplitude (\ref{Eq:cADP}) is rewritten as
\bea \label{Eq:cADP2}
\cA^{(\alpha)}_{\cI,\cJ} = \sum_{d\in D_{\cI, \cJ}} \sum_{\pi \in d} w(\pi) \, .
\eea
The point now is that there is only an exponential number of diagrams in $D_{\cI, \cJ}$ with a superexponential number of paths in each diagram, but that due to the strong correlations between the $w(\pi)$ in a given diagram the sum $ \sum_{\pi \in d} w(\pi)$ might be recast into a sum over only a smaller number of amplitudes associated with `effective paths':
\bea
\sum_{\pi \in d} w(\pi) = \sum_{\Gamma \in \cP(d)} \tilde{w}(\Gamma) \, ,
\eea
where the statistical properties of $ \tilde{w}(\Gamma)$ are essentially equivalent to those of $w(\pi)$. The key point is that the number of diagrams is only exponential in $N$ and that the number of effective paths per diagram $\cP(d)$ is also exponential in $N$. This is argued to hold in general in \cite{RosMullerScardicchio2015}. Here we only illustrate this notion of effective paths on an explicit example.

\paragraph{Example with $3$ vertices} We consider the diagram $d$ that is made of the three paths represented in Fig.~\ref{Fig:effectivepaths}, all connecting $(\alpha,\alpha)$ to $(\cI,\cJ)$ with $\cI=(\beta_1,\beta_2,\alpha_2,\alpha_3)$ and $\cJ=(\gamma_2,\gamma_1,\delta_3,\gamma_3)$ with one intermediate particle excitation $\alpha_1$. The different paths amplitudes $w(\pi)$ only differ from one another by the energy denominators that are encountered (the interaction terms $\eta$ and the overall minus sign $(-1)^{\sigma_{{\rm path}}}$) is identical. The sum over the three paths amplitudes is thus rewritten schematically
\bea
\sum_{\pi \in d} \frac{w(\pi)}{\lambda^2 \eta_1\eta_2\eta_3} & \sim &   \underbrace{\frac{1}{\cE_3(\cE_3+\cE_1)(\cE_3+\cE_1+\cE_2)}}_{a}  + \underbrace{\frac{1}{\cE_1(\cE_1+\cE_3)(\cE_1+\cE_3+\cE_2)}}_{b}  +\underbrace{\frac{1}{\cE_1(\cE_1+\cE_2)(\cE_1+\cE_2+\cE_3)}}_{c} \nn \\
&\sim &  \frac{1}{\cE_3 \cE_1(\cE_1+\cE_2)} \, , \nn 
\eea
with $\cE_1 := E_{\alpha_1}+E_{\beta_1}-E_{\gamma_1}-E_{\alpha}$, $\cE_2:=E_{\beta_2}+E_{\alpha_2}-E_{\gamma_2}-E_{\alpha_1}$, $\cE_3:=E_{\alpha_3}+E_{\alpha}-E_{\delta_3}-E_{\gamma_3}$. Hence here the sum over the three paths weight is equivalent to a single effective-path weight
\bea \label{Eq:EffectivePathWeightExample}
\sum_{\pi \in d} w(\pi) = \tilde{w}(\Gamma) \quad , \quad |\tilde{w}(\Gamma) | \sim \frac{\lambda^2 \eta_1\eta_2}{\cE_1(\cE_1+\cE_2)}   \frac{\lambda \eta_3}{\cE_3 } \, .
\eea
Here the fact that the three diagrams only corresponds to one effective path is due to the fact that the particle $c_{\alpha}^{\dagger}$ and the hole $c_{\alpha}$ independently decay into $c_{\beta_1}^{\dagger} c_{\beta_2}^{\dagger} c_{\alpha_2}^{\dagger} c_{\gamma_2}c_{\gamma_1}$ and $c_{\alpha_3}^{\dagger}c_{\delta_3} c_{\gamma_3}$ and (\ref{Eq:EffectivePathWeightExample}) is just the product of the amplitudes associated with these two decay processes. For diagrams with more complex branching structure the sum over path reduces to a sum over several effective paths, and the number of these effective paths is exponential in $N$.

\subsection{Large deviation of path weight} \label{subsec:MBL2:LD}

In order to check the MBL criterion (\ref{Eq:ExpansionLIOMsConvergenceN}) we need to estimate the large deviation function of effective paths weights $\phi(x)$ defined by  $\mathbb{P}(\frac{\log |\tilde{w}(\Gamma)|}{N} = x) \sim_{N \gg 1} e^{-N \phi(x)}$ (to leading order). As a model example of such effective paths weights $|\tilde{w}(\Gamma)|$ of length $N$ we consider
\bea
\tilde{w}_N = \prod_{i=1}^{N-1} \frac{\lambda |\eta_i|}{|s_i|} \,, \nn 
\eea
with the sequence of random energies $s_i$ given by $s_i = \sum_{j=1}^i \cE_j /\delta_{\xi} $ with the $\cE_j/\delta_{\xi}$
 some iid centered unit Gaussian random variables and the $\eta_i$ iid uniformly distributed random-variables in $[-1,1]$. In principle the numerator could also contribute to large deviations since for $x'\geq 0$, $\mathbb{P}(\sum_{i=1}^N \frac{\log |\eta_i|}{N} = -x') = \int_{u \in \JR} \frac{du}{2\pi} e^{+iux' - N \log(1+iu/N)} \sim_{N \gg 1} e^{-N(x'-1-\log(x'))}$. These large deviations of the numerator do not however control the position of the MBL critical point (which is controlled by the probability of having very large $\tilde{w}_N$, i.e. small denominators) in the perturbative regime $\lambda \ll 1$ and we replace the denominator by its typical value $ \prod_{i=1}^{N-1} \lambda |\eta_i| \sim (\lambda \eta_{{\rm typ}} )^{N}$ with $\eta_{\rm typ} =e^{-1}$. On the other hand it can be shown using a transfer matrix calculation (i.e. the recursive structure of the denominators) \cite{RosMullerScardicchio2015} that 
 \bea
\mathbb{P}\left(  \frac{1}{N} \log\left(  \prod_{i=1}^{N-1} \frac{1}{|s_i|}   \right) = \tilde{y} \right) \sim_{N \gg 1} \left(\frac{2e}{\sqrt{2\pi}} \tilde{y} \right)^N e^{-N {\cal F}(\tilde{y})} \, .
 \eea
With a large deviation function that has the asymptotic behavior (the relevant regime for $\lambda \ll 1$)
\bea \label{Eq:TailLargeDeviationCorrelated}
{\cal F}(\tilde{y}) =_{\tilde{y} \gg 1} \tilde{y} - \frac{\gamma_{E}}{2 \tilde{y}} + O(1/\tilde{y}^2) \, .
\eea
Interestingly the main (linear) behavior of the tail of the large deviation function is identical to the one that is obtained when taking the $s_i$ as iid unit centered Gaussian random variables. Indeed, in that case one easily obtains using Laplace transform
\bea
\mathbb{P}\left(  \frac{1}{N} \log\left(  \prod_{i=1}^{N-1} \frac{1}{|s_i|}   \right)  = \tilde{y} \right)  = \frac{1}{2 i \pi} \int_{B} e^{N(\tilde{y} u + \log g(u))} \nn 
\eea
with $g(u) = \frac{2^{\frac{1+k}{2}} \Gamma(\frac{1+k}{2})}{\sqrt{2 \pi}} $ and $B$ a contour passing to the right of the poles of the integrand. This integral can be evaluated using a saddle-point calculation and to leading order $\mathbb{P}\left(  \frac{1}{N} \log\left(  \prod_{i=1}^{N-1} \frac{1}{|s_i|}   \right)  = \tilde{y} \right) \sim e^{N ( u^*(\tilde{y}) + \log g(u^*(\tilde{y})))}$ with $\tilde{y}= - \frac{1}{2} \log(2) - \frac{1}{2} \psi(\frac{1+u^*}{2})$ with $\psi(x)=\Gamma'(x)/\Gamma(x)$ the diGamma function. The position of the saddle-point is for $\tilde{y} \gg 1$ at leading order given by $u^*=-1+2/\tilde{y} + o(1/\tilde{y})$ (this uses $\psi(z) =_{z \gg 1} -1/z = o(1/z)$). One obtains $\mathbb{P}\left(  \frac{1}{N} \log\left(  \prod_{i=1}^{N-1} \frac{1}{|s_i|}   \right) \right) \sim e^{N\left( -\tilde{y} + \log(\tilde{y}) - \frac{1}{2} \log(\frac{\pi}{2e^2}) + \frac{\log(2)-\gamma_E}{2y}  \right)} $. The tail of the large deviation function is thus identical to (\ref{Eq:TailLargeDeviationCorrelated}) up to a factor $\frac{\log(2)}{2 \tilde{y}}$ that enhances the probability to observe a large deviation. This is because in both the correlated and uncorrelated cases, the tail of the large deviation function is dominated by the same events where all the $\cE_i$ are small simultaneously.

Including the numerator in the effective-path weight we thus obtain that the PDF of large deviation of the path weights is to leading order given by
\bea \label{Eq:LDPMBL}
\mathbb{P}\left(\frac{\log |\tilde{w}(\Gamma)|}{N} = \tilde{x} + \log(\lambda \eta_{{\rm typ}})\right) \sim_{N \gg 1} \left(\frac{2e}{\sqrt{2\pi}} \tilde{x} \right)^N e^{-N {\cal F}(\tilde{x})} \, .
\eea

\subsection{Estimation of the number of diagrams} \label{subsec:MBL2:NumbOfDiagrams}

The large deviation of the effective-path weight is one of the ingredient needed to check the MBL criterion (\ref{Eq:ExpansionLIOMsConvergenceN}). As in the FWA treatment of Anderson localization on the Bethe lattice, we also need an estimation of the exponential growth of the total number of effective paths at the level $N$:
\bea
{\cal N}_N:= \sum_{\cI\neq \cJ , \, |\cI|=|\cJ|=N +1}  \, \sum_{d \in \cD_{\cI,\cJ}} \sum_{\Gamma \in \cP(d)} 1 \, . \nn 
\eea

\paragraph{Total number of trees at the level $N$} We first compute exactly the number of possible diagram {\it geometry} at the level $N$, $\cT_N$. Because of the peculiar geometry of the diagrams (see Fig.~\ref{Fig:diagrams1}) we have
\bea
\cT_N = \sum_{n_1,n_2 \geq 0, \, n_1+n_2= N} T_{n_1} T_{n_2} \, ,  \nn 
\eea
where $T_n$ is the number of trees with $n$ vertices (including the root) and branching ratio $3$. The latter satisfies the recursion
\bea
T_0=1 \quad , \quad T_{n+1} = \sum_{n_1+n_2+n_2 = n } T_{n_1} T_{n_2} T_{n_3} \, . \nn 
\eea
This implies that the generating function $T(x) := \sum_{n \geq 0} T_n x^n$ satisfies the functional equation
\bea
T(x) = 1 + x T(x)^3 \, . \nn 
\eea
From this, the full sequence $T_n$ can be obtained using Lagrange inversion theorem with $x(T) = \frac{T-1}{T^3}$
\bea
T_n = \frac{1}{n(n-1)!} \frac{d^{n-1}}{dT^{n-1}} \left(  \frac{T-1}{x(T)} \right)^n |_{T=1} = \frac{1}{2n+1} C^{n}_{3n} \,, \nn 
\eea
with $C^m_n$ the binomial coefficient. From this one gets, rewriting the factorial as Gamma function and using a resummation formula
\bea
\cT_n = \frac{3^{3/2+3n}}{\pi} \frac{\Gamma(n+2/3) \Gamma(n+4/3)}{\Gamma(2n+3)} \sim \frac{3}{4} \sqrt{\frac{3}{\pi}} \frac{1}{n^{3/2}} \left( \frac{27}{4}  \right)^n \, . \nn 
\eea

\paragraph{Upper bound for the number of diagrams} ${\cal T}_N$ gives the total number of possible diagram geometries at level $N$. Since any diagram is specified by a geometry and a labeling of the vertices by admissible (i.e. permitted by the interaction) single orbital indices, an upper-bound on the number of diagram is easily obtained by noting that the number of possible labeling per vertex is at most $3 {\cal K}$ (see Sec.~\ref{subsec:MBL2:FWA}). From this we get that the total number of diagrams is bounded (at the exponential level of accuracy) by
\bea
\sum_{\cI\neq \cJ , \,  |\cI|=|\cJ|=N +1} \, \sum_{d \in \cD_{\cI,\cJ}}  1 \leq \left( \frac{27}{4}  \right)^N (3 {\cal K} )^N  \sim (20.25\ {\cal K})^N \, . \nn 
\eea
This is an upper bound since it does not take into account effects such as Pauli's exclusion principle.

\paragraph{Lower bound for the number of diagrams}

A lower-bound on the total number of diagrams can also be obtained by considering the subset of diagrams with a so-called `necklace' geometry, which permits to obtain 
\bea
\sum_{\cI\neq \cJ , \,  |\cI|=|\cJ|=N +1} \, \sum_{d \in \cD_{\cI,\cJ}}  1  \geq (10.6\ {\cal K})^N \, . \nn 
\eea 
We refer to \cite{RosMullerScardicchio2015} for details but note that this lower-bound is not important to show localization, but permits to refine the estimation of the MBL transition point in the FWA (the simplest lower-bound would be to take the total number of trees $(27/4)^N$), the crucial point being to be able to show that the total number of effective paths at the level $N$ grows only exponentially with $N$.

\paragraph{Number of effective paths per diagram and final estimate} The final ingredient is to obtain a bound on the total number of effective paths per diagram $ |{\cal P}(d)| := \sum_{\Gamma \in \cP(d)} 1  $. That can also be done by considering a subclass of diagram and one obtain the bound (see \cite{RosPHD} for details) ${\cal P}(d) \leq |\overline{\cal P}(d)| \simeq e^{0.58 N}$. Combining everything one finds that (to exponential accuracy)
\bea \label{Eq:TotalNumberOfDiagrams}
{\cal N}_N \simeq (C {\cal K})^N \, 
\eea
with $18.97<C<36.25$

\subsection{Estimation of the MBL transition point} \label{subsec:MBL2:Estimate}

We now estimate the position of the MBL transition point using the MBL criterion (\ref{Eq:ExpansionLIOMsConvergenceN4}). We thus compute, assuming that we are in the localized phase and that this probability converges to $1$,
\bea
\mathbb{P}\left( {\rm max}_{\cI\neq \cJ , |\cI|=|\cJ|=N +1}  |\cA_{\cI,\cJ}^{(\alpha)}|  < z^N\right)  &=& \prod_{\cI\neq \cJ , |\cI|=|\cJ|=N +1} \left( 1 - \mathbb{P}\left( |\cA_{\cI,\cJ}^{(\alpha)}|  > z^N\right)  \right) \nn \\
&\simeq& \exp\left[ -  \sum_{\cI\neq \cJ , |\cI|=|\cJ|=N +1}   \mathbb{P}\left( |\cA_{\cI,\cJ}^{(\alpha)}|  > z^N\right)  \right]\nn
\eea
Now, since $\cA_{\cI,\cJ}$ is itself the sum of $|\cD_{\cI,\cJ}| \overline{\cP(d)} $ effective paths, which are also distributed with a fat tail distribution, we can approximate, assuming that different effective paths are uncorrelated and in the regime where the computed probability is small (relevant in the localized phase),
\bea
\mathbb{P}\left( |\cA_{\cI,\cJ}^{(\alpha)}|  > z^N\right)  \simeq |\cD_{\cI,\cJ}| \overline{\cP(d)} \mathbb{P}\left( |\tilde{w}(\Gamma)|  > z^N\right) \nn  \, .
\eea
Putting everything together, we obtain that in the localized phase we must have
\bea
1 =  \lim_{N \to \infty} \mathbb{P}\left( {\rm max}_{\cI\neq \cJ , |\cI|=|\cJ|=N +1}  |\cA_{\cI,\cJ}^{(\alpha)}|  < z^N\right)  = \lim_{N \to \infty} \exp\left( -  {\cal N}_N \mathbb{P}\left( |\tilde{w}(\Gamma)| > z^N\right)  \right) \, . \nn 
\eea
Using  \eqref{Eq:LDPMBL}  we estimate
\bea
\mathbb{P}\left( |\tilde{w}(\Gamma)| > z^N\right)  && \simeq \left( \frac{2 e}{\sqrt{2\pi}} \right)^N \int_{ \log(\frac{z}{\lambda \eta_{{\rm typ}}})} \tilde{x}^N e^{-N {\cal F}(\tilde x)} d \tilde{x}   \nn \\
&& \simeq \left( \frac{2 e}{\sqrt{2\pi}} \log(\frac{z}{\lambda \eta_{{\rm typ}}}) \right)^N e^{-N  \log(\frac{z}{\lambda \eta_{{\rm typ}}})} \nn 
\eea

Using \eqref{Eq:TotalNumberOfDiagrams} we thus get the condition that, in the localized phase, there exist $z\in ]0,1[$ such that
\bea
 \lim_{N \to \infty} \exp\left( -  e^{ N \log {\cal G}(\lambda , {\cal K} , z)}\right)  = 1 \, , 
\eea
with
\bea
{\cal G}(\lambda , {\cal K},z) := C {\cal K}  \frac{2 e}{\sqrt{2\pi}}  \frac{\lambda \eta_{{\rm typ}}}{z}  \log\left( \frac{z}{\lambda \eta_{\rm typ}}  \right) \, .
\eea
The position of the critical point $\lambda = \lambda_c$ is then obtained setting ${\cal G}(\lambda_c , {\cal K},z=1)=1$. In the limit of large ${\cal K}$ and using $\eta_{{\rm typ}}=1/e$ one gets 
\bea \label{Eq:ThresholdMBLFWA}
\lambda_c \simeq \frac{\sqrt{2\pi}}{2e C} \frac{1}{{\cal K} \log({\cal K})} \, .
\eea
I.e. a result that is formally equivalent to the one obtained using the FWA on the Bethe lattice \eqref{Eq:ResBetheLatticeWcFWA} if one replaces $W/g \to \lambda$ and $K \to (C/\sqrt{2\pi}) {\cal K}$.

\paragraph{Fermi blockade} The condition $\lambda < \lambda_c$ ensures at the FWA level the convergence of the expansion of the LIOMs \eqref{Eq:ExpansionLIOMsI}, which implies localization in operator space, and hence localization of all many-body eigenstates of the interacting Hamiltonian \eqref{Eq:MBHamRecall}. If one is only interested in localization of many-body eigenstates at a given electron density $\nu$, it is tempting to replace the connectivity ${\cal K}$ of the operator lattice by an effective connectivity ${\cal K}_{\rm eff} = \nu(1-\nu){\cal K}$. Indeed, among the terms appearing in the expansion \eqref{Eq:ExpansionLIOMsI} at order $N$, only a fraction of order $(\nu(1-\nu))^N$ do not annihilate a typical state, and the connectivity of the operator lattice that is relevant is thus reduced. Inserting into \eqref{Eq:ThresholdMBLFWA} leads to a dependence of the critical interaction strength on the electron density $\nu$. A similar argument can be formulated to take into account a finite temperature (i.e. looking at states at a given energy density), leading to many-body mobility edges in analogy with the single particle problem. This argument exists in BAA \cite{BAA2006}, and is correct at the level of the family of diagrams analyzed (leading order $1/M$ where the parameter $M$ is introduced in BAA eq.\ (48h)). Several researchers \cite{DeRoeckHuveneersMullerSchiulaz2016} deem this result as controversial, and pose the scenario in which the whole spectrum must be either localized or delocalized due to effects which are non-perturbative in the interaction (and in $1/M$). We do not have time to enter in this fiend now, nor anything new is gained by considering the construction of LIOMs. Furthermore, as argued in \cite{chandran2016many}, in higher dimensions $d$ LIOMs perturbation theory can be made divergent by \emph{boundary} effects, and a more careful definition of LIOMs is needed.

\paragraph{Funding information}
A.S. kindly acknowledges financial support from a 2017 Google Faculty Award. T.T. has been supported by the InterUniversity Attraction Pole phase VII/18 dynamics, geometry and statistical physics of the Belgian Science Policy. T.T. is a postdoctoral fellow of the Research Foundation-Flanders (FWO).

\bibliography{Biblio.bib}

\nolinenumbers

\end{document}